\documentclass[aps,twocolumn,prl,superscriptaddress,amsmath,amssymb,tightenlines,nofootinbib]{revtex4-2}

\usepackage{epsfig,graphicx,times}
\usepackage{booktabs}
\usepackage{amstext}
\usepackage{amsmath}            
\usepackage{amssymb}            
\usepackage{graphicx}           
\usepackage{latexsym}
\usepackage{bm}
\usepackage{color}
\usepackage {float}
\RequirePackage{jabbrv}
\makeatletter
\@ifundefined{textcolor}{}
{\definecolor{BLACK}{gray}{0}
 \definecolor{WHITE}{gray}{1}
 \definecolor{RED}{rgb}{1,0,0}
 \definecolor{GREEN}{rgb}{0,1,0}
 \definecolor{BLUE}{rgb}{0,0,1}
 \definecolor{CYAN}{cmyk}{1,0,0,0}
 \definecolor{MAGENTA}{cmyk}{0,1,0,0}
 \definecolor{YELLOW}{cmyk}{0,0,1,0}
}

\makeatother

\begin{document}

\title{Degeneracy-breaking and Long-lived Multimode Microwave Electromechanical Systems Enabled by Cubic Silicon-Carbide Membrane Crystals}

\author{Yulong Liu}
\email{liuyl@baqis.ac.cn}
\affiliation{Beijing Academy of Quantum Information Sciences, Beijing 100193, China}
\affiliation{Department of Applied Physics, Aalto University, P.O. Box 15100, FI-00076 Aalto, Finland}

\author{Huanying Sun}
\affiliation{Beijing Academy of Quantum Information Sciences, Beijing 100193, China}

\author{Qichun Liu}
\affiliation{Beijing Academy of Quantum Information Sciences, Beijing 100193, China}

\author{Haihua Wu}
\affiliation{Beijing Academy of Quantum Information Sciences, Beijing 100193, China}

\author{Mika A. Sillanp\"{a}\"{a}}
\affiliation{Department of Applied Physics, Aalto University, P.O. Box 15100, FI-00076 Aalto, Finland}
\date{\today}

\author{Tiefu Li}
\email{litf@tsinghua.edu.cn}
\affiliation{School of Integrated Circuits and Frontier Science Center for Quantum Information, Tsinghua University, Beijing 100084, China}

\begin{abstract}
Cubic silicon-carbide crystals (3C-SiC), known for their high thermal conductivity and in-plane stress, hold significant promise for the development of high-quality ($Q$) mechanical oscillators. We reveal degeneracy-breaking phenomena in 3C-phase crystalline silicon-carbide membrane and present high-$Q$ mechanical modes in pairs or clusters. The 3C-SiC material demonstrates excellent microwave compatibility with superconducting circuits. Thus, we can establish a coherent electromechanical interface, enabling precise control over 21 high-$Q$ mechanical modes from a single 3C-SiC square membrane. Benefiting from extremely high mechanical frequency stability, this interface enables tunable light slowing with group delays extending up to an impressive duration of \emph{an hour}. Coherent energy transfer between distinct mechanical modes are also presented. In this work, the studied 3C-SiC membrane crystal with their significant properties of multiple acoustic modes and high-quality factors, provide unique opportunities for the encoding, storage, and transmission of quantum information via bosonic phonon channels.
\end{abstract}
\maketitle

\noindent\textbf{Introduction}\\
Silicon carbide (SiC) is a compound semiconductor that manifests in multiple polytypes, each characterized by distinct crystal structures and stacking sequences of silicon (Si) and carbon (C) atoms. The distinctive advantages offered by individual SiC crystal structures play a pivotal role in tailoring SiC-based devices to specific applications and optimizing their performance~\cite{gerhardt2011properties,feng2023handbook,cheung2006silicon}. The hexagonal structure, such as 4H-SiC and 6H-SiC, represents the most commercially available high-quality SiC crystal form and has garnered widespread attention in the field of quantum information processing due to its potential for hosting quantum bits (qubits)~\cite{aharonovich2014Defects,boretti2014Siliconaspects,castelletto2022silicongates,lukin2020integrated,majety2022quantum,awschalom2018quantum,elshaari2020hybrid,lohrmann2017review,castelletto2020silicon,gao2015coherent,luo2023recent,luo2023fabrication,Wang2023Magnetic}. In contrast to hexagonal silicon carbide, cubic silicon carbide (3C-SiC) possesses a zinc-blende crystal structure that closely matches the lattice constant of silicon. Currently, 3C-SiC can be grown directly on a silicon substrate using heteroepitaxy techniques~\cite{zhuang2015electrochemical}. This compatibility facilitates the integration of 3C-SiC with existing microfabrication and machining technologies on silicon substrates~\cite{dao2016piezoresistive,phan2017single,chatterjee2021semiconductor}, which is challenging to achieve with bulk substrates like 4H-SiC and 6H-SiC. In recent years, tensile-loaded thin-film mechanical resonators have achieved significant attention due to their remarkably high $Q$-factors~\cite{yuan2015large,liu2023coherent,serra2021silicon,hoj2021ultra,tsaturyan2017ultracoherent,galinskiy2020phonon,sementilli2022nanomechanical}. Crystalline thin films, in general, have a higher theoretical limit for holding tensile stress compared to non-crystalline or amorphous thin membranes~\cite{beccari2022strained,shin2022spiderweb}. The stronger atomic bonding in 3C-SiC contributes to its higher ultimate tensile strength (approaching 18~GPa) compared to hexagonal polytypes~\cite{kwon2015room}. On the other hand, the heteroepitaxial growth techniques can also result in low gradient and high in-plane stress in 3C-SiC thin films~\cite{phan2017ultra}. The magnitude of in-plane stress can be controlled by adjusting the growth conditions including temperature, pressure, and growth rate~\cite{anzalone2014evaluation,rohmfeld2002quantitative,capano2006residual,colston2016mapping}. With these significant advantages, tensile-loaded 3C-SiC is emerging as the material of choice for developing mechanical resonators with high $Q$-factors~\cite{xu2023Amorphous}. It is important to note that, while 3C-SiC has a high theoretical upper limit of stress, the actual quality factors ($Q$-factors) of mechanical oscillators made from 3C-SiC are often limited by various factors such as imperfections in the crystal structure~\cite{kermany2016factors}. To date, 3C-SiC mechanical microresonator with doubly clamped string~\cite{henry2003nanodevice,kermany2014microresonators}, cantilever~\cite{zorman2008micro}, trampoline~\cite{romero2020engineering}, phononic crystal~\cite{anufriev2022nanoscale}, and suspended square membrane~\cite{barnes2012pressure,zhou2008fracture,nguyen2017superior} structures have been developed and measured in experiment. For the crystalline SiC mechanical resonator, the experimentally achieved highest quality factor reaches around one million ($Q \sim 10^{6}$) in high vacuum~\cite{kermany2014microresonators,zorman2008micro,romero2020engineering}.
\begin{figure*}[pth]
\includegraphics[scale=0.4]{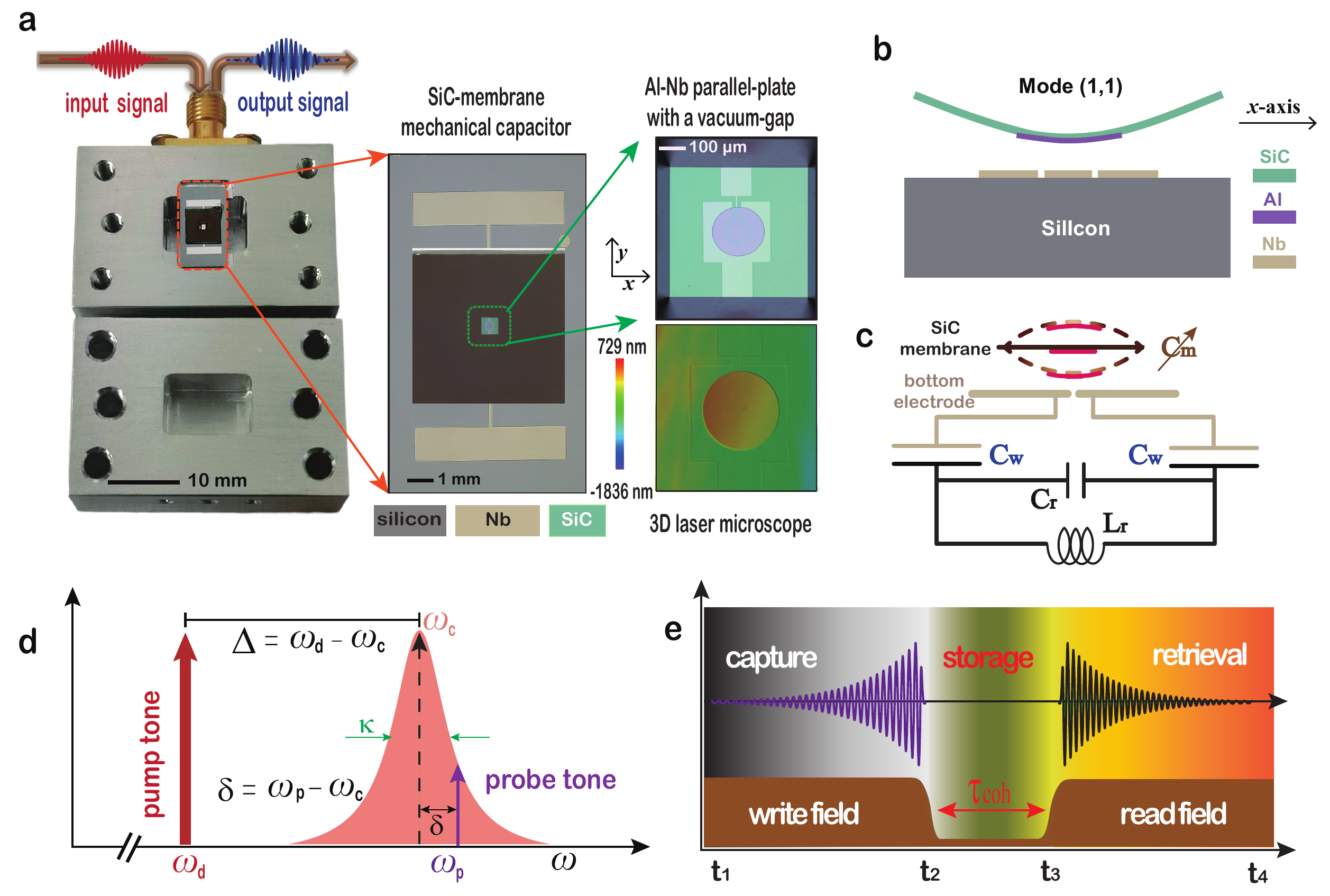}
\caption{\noindent\textbf{The microwave cavity electromechanical interface device and pump-probe schemes used in experiments.} \textbf{a} The superconducting 3D-cavity with a SiC-membrane-based mechanical capacitor chip inserted inside its strongest electric field concentration area, forming the dispersive type electromechanical interface device. The zoomed-in area shows the detailed structure of the mechanical parallel-plate capacitor formed by flipping a metalized SiC membrane on top of another bottom-electrode microwave antenna. With the geometric center of the bottom electrode as the origin, a notch appears upward in the $y$-axis direction. The green arrow marks the core area of the mechanical capacitor, viz., the upper and lower parallel electrode plates and the vacuum interlayer. 3D laser microscope of the parallel plate electrode region indicates the vacuum gap distance in capacitors. \textbf{b} A schematic illustrating the stacked arrangement of the different layers is presented, with a cut along the \textit{x}-axis. \textbf{c} The effective lumped-element circuit model for our device, where the photon radiation force causes SiC-membrane's mechanical vibration which in turn introduces a tunable capacitor ($C_{\textrm{m}}$) and changes the total cavity capacitance. \textbf{d} is the continuous wave pump-probe scheme. Strong pump-tone (marked by red-arrow) is detuned by $\omega_{\textrm{m}}$ from the cavity resonance frequency and weak probe-tone (marked by purple-arrow) is scanning the dynamics of interest occurred over the cavity mode, which has a linewidth of $\kappa$. \textbf{e} shows the time sequence of the pulsed pump-probe scheme which allows an on-demand writing, storage and reading of microwave pulses by using the long-lived phonons as the core storage element.}
\label{fig1}
\end{figure*}

In the effort to push mechanical oscillators into the quantum realm, the radiation-pressure interaction in laser optomechanical and microwave electromechanical systems plays a crucial role in, e.g., precisely manipulating the macroscopic quantum state of mechanical resonator~\cite{ockeloen2018stabilized,riedinger2018remote,suh2014mechanically,mercier2021quantum}, while also providing a transducer interface for connecting phonons with other quantum systems~\cite{kurizki2015quantum,han2021microwave,clerk2020hybrid}. In pursuit of one of the long-standing goals of fundamental tests of quantum mechanics, such as macroscopic-scale Bell tests~\cite{vivoli2016proposal,hofer2016proposal,marinkovic2018optomechanical}, quantum teleportation~\cite{fiaschi2021optomechanical}, and tests of quantum gravity~\cite{belenchia2017tests,gely2021superconducting,liu2021gravitational}, long-lived mechanical oscillators with extremely low thermal decoherence and pure dephasing rates are highly desired~\cite{cleland2024studying}. These urgent needs also arise in the field of quantum information processing, particularly in developing microwave-light interfaces for quantum computation and communication~\cite{sahu2022quantum,arnold2020converting,weaver2024integrated,honl2022microwave,stockill2022ultra,shen2024photonic,delaney2022superconducting,mirhosseini2020superconducting}. Strain engineering and dissipation dilution with trampoline~\cite{kleckner2011optomechanical,barzanjeh2019stationary,manjeshwar2023high}, nested~\cite{weaver2016nested,Weaver2017Coherent,bereyhi2022hierarchical,PhysRevLett.124.025502}, and acoustic isolation structures~\cite{brubaker2022optomechanical,maccabe2020nano,kristensen2024long,fedoseev2021stimulated,planz2023membrane} have been extensively explored in laser and microwave optomechanical devices to mitigate extrinsic mechanical dissipation. While acoustic radiation into the substrate has been significantly suppressed due to weak connecting links to the substrate, this comes at the cost of reduced thermal conductivity to the environment~\cite{huang2024room,saarinen2023laser,norte2016mechanical,reinhardt2016ultralow}. In addition, the carefully designed mechanical structure for dissipative dilution poses significant challenges for the construction of microwave cavity optomechanical devices~\cite{beccari2022strained,ghadimi2018elastic,PhysRevX.12.021036,reetz2019analysis,shaniv2023understanding}. The reduced thermal conductivity, combined with an increase in two-level systems or defects, introduces additional mechanical damping and further diminishes frequency stability~\cite{kalaee2019quantum,fink2016quantum,barzanjeh2017mechanical,seis2022ground}. These factors present substantial obstacles for engineered high-$Q$ mechanical resonators in achieving lower thermal decoherence and pure dephasing rates in optomechanical devices~\cite{bozkurt2023quantum}. Using single-crystal silicon oscillators as an example, phonon bandgap engineering has significantly lowered the energy relaxation rate, achieving an energy quality factor as high as \(10^{10}\)~\cite{beccari2022strained}. When these oscillators are combined with optical or microwave resonant cavities, the phonon lifetime remains high at 1.5 seconds. However, temperature-dependent frequency jitter and residual two-level system defects reduce the final coherence time of the vibration mode to the microsecond level, representing a decrease of over four orders of magnitude~\cite{maccabe2020nano,bozkurt2023quantum}. Therefore, developing mechanical oscillators with both high thermal conductivity and high quality factor is crucial, as this can enhance frequency stability and increase coherence time. Recently, both theoretical calculations from first principles~\cite{meng2019phonon} and experimental measurements at the wafer-scale~\cite{Cheng2022Highthermalconductivity} have revealed that 3C-SiC films exhibit isotropic high thermal conductivity at room temperature, which is the second highest among types of crystals and only surpassed by diamond~\cite{yan2021single}. In the field of quantum information processing, controlling and dissipating heat generated by quantum devices are crucial for maintaining quantum coherence~\cite{maillet2020electric}.

In this work, the high thermal conductivity of 3C-SiC mechanical resonators at low temperatures ensures effective thermalization to millikelvin levels, resulting in exceptional frequency stability and extremely small pure-dephasing. We highlight the distinctive non-uniform tensile stress found in crystalline membranes and provide insights into the multiple degeneracy-broken mechanical mode pairs and clusters. Furthermore, we explore their potential as numerous high-$Q$ and long-lived phononic modes for quantum information processing, particularly focusing on their extremely low pure dephasing rates and high frequency stability in conjunction with superconducting quantum circuits. Benefiting from extremely high mechanical frequency stability, this electromechanical interface enables tunable light slowing with group delays extending up to an impressive duration of \emph{an hour}. While numerous experimental protocols and theoretical studies have aimed to enhance the group delay of light, our work achieves a substantial improvement, offering several orders of magnitude advancement over previous pioneering efforts~\cite{kristensen2024long,fink2016quantum,teufel2011circuit,zhou2013slowing,lake2021processing,shandilya2021optomechanical,merklein2017chip,JingHui2023Magnon,YingWu2019Bose}. We systematically explore on-demand state storage and energy transfer between degeneracy-broken high-$Q$ mechanical modes, achieving quality factors exceeding \(10^{8}\). Enabling coherent electromechanical manipulation between long-lived multiple mechanical resonators would be instrumental in advancing the development of phononic memories, repeaters, and transducers for microwave quantum states.
\bigskip
\\
\noindent\textbf{Results}\\
\noindent\textbf{3C-SiC based microwave electromechanical system}\\
As shown in Fig.~1\textbf{a}, our cavity electromechanical device consists of a three-dimensional (3D)  superconducting microwave resonator and a mechanical parallel-plate capacitor chip. The 3D microwave cavity is made from bulk aluminum (Al-6061) with a rectangle recess ($8\times12\times16~$mm). The panel at the right of the device shows the detailed structure of the mechanical capacitor which is constructed by flipping the metalized 3C-SiC membrane chip on top of another microwave coupling chip with antenna electrodes on its front surface. The SiC membranes are 50~nm thick grown on 400~$\mu$m thick, width 5$\times$5~mm square silicon frames (with a low resistivity). The measured X-ray diffraction pattern indicates that the grown SiC thin-film is in 3C phase. The suspended membrane window is a square measuring 500~$\mu$m$\times$500~$\mu$m, with the backside silicon of the membrane removed through potassium hydroxide wet etching. The central region of the membrane window was deposited by a 20-nm thick aluminum (Al) film through an electron beam evaporator. The metalized electrode on the suspended 3C-SiC membrane working as the upper plate of the parallel-plate capacitor is circular with a diameter of 200~$\mu$m. The lower plate chip is fabricated by sputtering a thin layer of niobium (Nb) with a thickness of 120~nm on top of a high-resistivity silicon substrate ($\langle100\rangle$ oriented) with a thickness of 500~$\mu$m.

The zoomed-in subfigure in Fig.~1\textbf{a}, right to the capacitor chip, shows the core area of the mechanical capacitor. The bottom electrode structure, in shape of the letter ``H", serves as the lower plate of mechanical parallel-plate capacitor and also functions as a microwave antenna, wireless connecting the mechanical capacitor to the 3D microwave cavity resonator. The center of the membrane is metallized, making the mechanical capacitance changes particularly sensitive to modes with vertical displacement at the membrane's center. Additionally, the presence of a notch in the upper half along the \(y\)-axis of the bottom electrode facilitates the detection of higher-order vibration modes, particularly those exhibiting vertical displacement at the notch position. This will be discussed in detail in the following sections. An optical microscope photographed the entire parallel plate electrode area through the SiC membrane window. The green interference pattern indicates the gap between the SiC membrane and the bottom silicon substrate, while the purple interference pattern originates from interference of light inside the vacuum gap of the capacitor. To get a more accurate value, we use a 3D laser microscope to measure the height difference between the top and bottom electrodes. From the line-cut height profile (see Supplementary Note~1), it can be inferred that the vacuum gap is around 576~nm. The side view, cut along the \textit{x}-axis, schematic illustrates the stacked arrangement of the different layers, as presented in Fig.~1\textbf{b}.

The effective lumped-element circuit model of the device is shown in Fig.~1\textbf{c}, where the photon radiation pressure couples the microwave cavity field to the mechanical vibration of the SiC-membrane through a position-dependent capacitor with a capacitance of C$_{\textrm{m}}(\hat{x})$. The variable mechanical capacitance changes the cavity resonance frequency, forming dispersive type cavity optomechanical interaction. The interaction Hamiltonian can be described as $H_{\textrm{int}}^{i}/\hbar=g_{i}a^{\dag}a\hat{x_{i}}$. The coupling parameter is $g_{i}=\partial\omega_{c}/\partial x_{i}$, and $\hat{x_{i}}=x_{\textrm{zpf}}^{i}(b_{i}+b^{\dag}_{i})$ is the displacement operator for each distinct mechanical mode. The mode shape identifier $i=(k,l)$, with $k$ and $l\in \mathbb{N}^+$, refers to the number of humps (antinodes) in the $x$ and $y$ directions of our square 3C-SiC membrane. Here, $a^{\dag}$ ($a$) and $b_{i}^{\dag}$ ($b_{i}$) are quantized bosonic creation (annihilation) operators of microwave cavity field and the mechanical mode $(k,l)$, respectively. Here, $\omega_{c}$ is the cavity mode angular resonance frequency, $x_{\textrm{zpf}}^{i}=\sqrt{\hbar/(2m_{i}\omega_{\textrm{m}}^{i})}$ is the root-mean-square amplitude of each oscillator's zero-point fluctuations, where $m_{i}$ and $\omega_{\textrm{m}}^{i}$ represent the effective mass and the resonance frequency of the corresponding mechanical mode $(k,l)$. When considering photons in the microwave frequency domain, this system is referred to as a cavity electromechanical system. Due to the fact that these mechanical modes are well separated by their resonant frequency, The pump and probe schemes for each mechanical modes will not affect the other. Thus, in the following, we do not specify the identifier `$i$' unless it is absolutely necessary.

The Hamiltonian used to describe each mechanical mode that is electromechanically coupled to the microwave cavity field, thereby forming a standard cavity optomechanical system in the microwave regime, is expressed as follows: $H/\hbar=\omega_{\textrm{c}}a^{\dag}a+\omega_{\textrm{m}}b^{\dag}b+g_{0}a^{\dag}a\left(b+b^{\dag}\right)$, where $g_{0}=gx_{\textrm{zpf}}$ is the vacuum optomechanical coupling strength. We add a continuous driving tone with frequency $\Omega_{\textrm{c}}$ and amplitude $\xi$ to enhance the effective radiative pressure interaction. As shown in Fig.~1\textbf{d}, the frequency of driving-tone is red-detuned to the cavity mode resonant frequency ($\omega_{\textrm{c}}$) by a mechanical frequency ($\omega_{\textrm{m}}$), i.e., $\omega_{\textrm{d}}=\omega_{\textrm{c}}-\omega_{\textrm{m}}$. Working in the rotating frame at the pump-tone frequency, the linearized Hamiltonian is described as
\begin{equation}
H/\hbar=\omega_{\textrm{m}}\left(a^{\dag}a+b^{\dag}b\right)+G\left(a^{\dag}+a)(b+b^{\dag}\right). \label{SiCHamiltonian}
\end{equation}
Here, $G=g_{0}\sqrt{N_{\textrm{c}}}$ is the linearized and driving-field enhanced coupling strength. $N_{\textrm{c}}$ is the occupation number of photons in cavity mode. Thus, the coupling rate $G$ can be continuously tuned by controlling the power of external coherent driving-tone~\cite{aspelmeyer2014cavity}.
\begin{figure*}[ptb]
\centering
\includegraphics[scale=0.95]{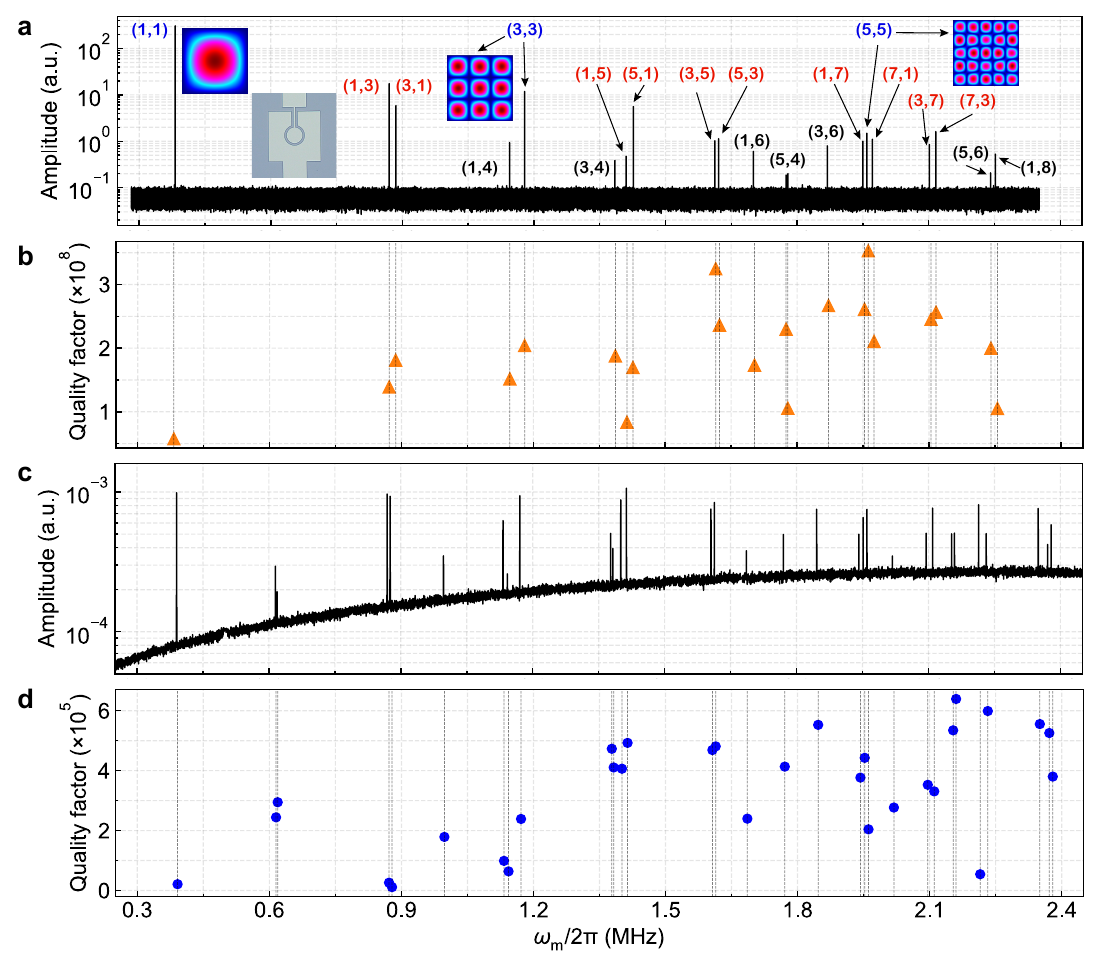}
\caption{\noindent\textbf{The microwave and optical readout of multiple mechanical modes from a single square 3C-SiC crystalline membrane.}~\textbf{a} The microwave power spectral density (PSD) readout of mechanical modes and the corresponding mode numbers ($k$,$l$) are labeled around each peak of the measured mechanical modes. The finite element method (FEM) simulations with non-uniform tensile stress for mode (1,1), (3,3) and (5,5) are presented. A notch structure extends upward along the \(y\)-axis of the bottom electrode. \textbf{b} The corresponding $Q$-factors for each mechanical mode in a dilution refrigerator. The $Q$-factors are obtained through the stroboscopic ringdown measurements. Twenty-first mechanical modes can be measured based on their independent optomechanical interaction with the same cavity field.~\textbf{c} The mechanical modes from the same crystalline 3C-SiC square membrane are detected with a laser Doppler vibrometer at room temperature and corresponding $Q$-factors are present in~\textbf{d}. The laser spot is focused near the center of the square membrane.}
\label{fig2}
\end{figure*}

Taking the cavity and mechanical decay rates into consideration, the set of Heisenberg-Langevin equations of a microwave electromechanical device in the presence of a driving tone are given as follows:
\begin{eqnarray}
\dot{a} &=&\left(i\Delta-\kappa/2\right) a-iG\left(b+b^{\dag}\right) + \sqrt{\kappa_{s}}\xi_{s},\label{Langevin1} \\
\dot{b} &=&-\left(i\omega_{\textrm{m}}+\gamma_{\textrm{m}}/2\right) b-iG\left(a+a^{\dag}\right) +\sqrt{\gamma_{\textrm{m}}}\xi_{\textrm{m}}, \label{Langevin2}
\end{eqnarray}
where $s=\left\{\textrm{in},\textrm{ex}\right\}$. Here, $\kappa_{\textrm{in}}$ and $\kappa_{\textrm{ex}}$ are the intrinsic and external microwave coupling rate to their dissipative bath, $\kappa$ is the total microwave loss rate. The symbol $\gamma_{\textrm{m}}$ represents the total decay rate of the studied mechanical mode of the square 3C-SiC membrane. The variable $\Delta$ denotes the frequency detuning between the driving pump tone and the resonant frequency of the cavity mode, defined as $\Delta = \omega_{\textrm{d}} - \omega_{\textrm{c}}$. The corresponding input noise operators satisfy the commutation relations $\langle\xi_{\textrm{in}}^{\dag}(t)\xi_{\textrm{in}}(0)\rangle=N_{c}^{\textrm{th}}\delta(t)$, $\langle\xi_{\textrm{in}}(t)\xi_{\textrm{in}}^{\dag}(0)\rangle=(N_{c}^{\textrm{th}}+1)\delta(t)$, and $\langle\xi_{o}^{\dag}(t)\xi_{o}(0)\rangle=N_{o}^{\textrm{th}}\delta(t)$, $\langle\xi_{o}(t)\xi_{o}^{\dag}(0)\rangle=(N_{o}^{\textrm{th}}+1)\delta(t)$. $N_{c}^{\textrm{th}}$ is the cavity mode thermal bath occupation. Subscripts $o=\left\{\textrm{ex},\textrm{m}\right\}$ represent the entering noise from cavity mode external pump phase noise $N_{\textrm{ex}}^{\textrm{th}}$, and mechanical thermal bath occupation $N_{\textrm{m}}^{\textrm{th}}$, respectively. From the input-output theory, the output field can be expressed as $a_{\textrm{out}}=\xi_{\textrm{ex}}-\sqrt{\kappa_{\textrm{ex}}}a$. Based on the Wiener-Khinchin theorem, we can now calculate the symmetrized noise power spectrum density (PSD) of the output microwave field~\cite{RevModPhys.82.1155}. Taking the Fourier transformation of the time derivative equations of Eqs.~(\ref{Langevin1})-(\ref{Langevin2}), and using the commutation relations of noise operators in the frequency domain, the PSD of output mode $S[\omega]$ can be expressed as
\begin{equation}
S[\omega]=\frac{1}{2}\left\langle a_{\mathrm{out}}^{\dag }\left[-\omega \right] a_{\mathrm{out}}\left[\omega\right] +a_{\mathrm{out}}\left[-\omega\right] a_{\mathrm{out}}^{\dag }\left[\omega\right]\right\rangle.  \label{frequencypsd}
\end{equation}

The optical-damping rate induced by the microwave drive dynamical backaction is $\gamma_{\textrm{opt}}=4G^{2}/\kappa$. In our experiment, the device always works in the weak optomechanical coupling regime ($G\ll\kappa$). When applying the maximum sideband pump power, the optomechanical damping rate will not exceed one percent of the microwave cavity linewidth (viz., $\gamma_{\textrm{opt}}\ll\kappa$). Under the aforementioned conditions, the PSD of the output cavity field is ultimately expressed as~\cite{hertzberg2010back}
\begin{equation}\label{PSDoutputfinal}
S[\omega]=n_{\textrm{add}}+\frac{4\kappa_{\textrm{ex}}\kappa N_{\textrm{c}}^{\textrm{th}}}{\kappa^{2}+4\tilde{\delta}^{2}}+\frac{4\eta\gamma_{\textrm{opt}}\gamma_{\textrm{tot}}\left(N_{\textrm{m}}^{\textrm{th}}-2N_{\textrm{c}}^{\textrm{th}}\right)}{\gamma_{\textrm{tot}}^2+4\tilde{\delta}^2},
\end{equation}
where $\tilde{\delta}=\omega-\omega_{m}$, and $\eta=\kappa_{\textrm{ex}}/\kappa$. The total mechanical damping rate is defined as $\gamma_{\textrm{tot}}=\gamma_{\textrm{m}}+\gamma_{\textrm{opt}}$.

During the microwave measurement, the added noise, $n_{\textrm{add}}$, mostly coming from the HEMT amplifiers and to a small extent from vacuum noise, creates the background noise floor of the measured PSD. The effective microwave thermal occupation $N_{\textrm{c}}^{\textrm{th}}$ may come from the thermal effects caused by defects in bulk superconducting cavities. These defects or impurities can induce thermal effects through ways of increasing surface resistance, causing localized heating, or promoting quasiparticle generations. $N_{\textrm{m}}^\textrm{th}$ is the effective thermal occupation of mechanical mode and transfers into the microwave output field via the electromechanical coupling. The second term in Eqs.~(\ref{PSDoutputfinal}) indicates that cavity thermal noise spectrum is a Lorentzian type with a linewidth of $\kappa$. The last term in Eqs.~(\ref{PSDoutputfinal}) indicates that the damped mechanical mode is featured as a Lorentzian spectrum with a total linewidth of $\gamma_{\textrm{tot}}$. The full-noise PSD contains two Lorentzian peaks, where the mechanical sidebands are superimposed on the Lorentzian peak of the cavity thermal emission.

When measuring the PSD spectrum, we select a driving power that allows the optomechanical damping and cavity thermal occupation to be negligible, i.e.,~$\gamma_{\textrm{opt}}\ll\gamma_{\textrm{m}}$, and $N_{\textrm{c}}^{\textrm{th}}\simeq0$. Under such conditions, the expression of the PSD spectrum reduces into a standard Lorentzian function
\begin{equation} \label{PSDdegrades}
\tilde{S}[\omega]=n_{\textrm{add}}+\mathcal{G}\frac{\gamma_{\textrm{m}}/2}{\left(\gamma_{\textrm{m}}/2\right)^{2}+\left(\omega -\omega_{\mathrm{m}}\right)^{2}},
\end{equation}
where $\mathcal{G}=2\eta \gamma_{\mathrm{opt}}N_{\textrm{m}}^{\text{th}}$.

In the following sections, we will first employ the Power Spectral Density (PSD) method to identify all the mechanical modes detectable by the microwave cavity. We will particularly focus on how the breaking of degeneracy generates multiple detectable high-$Q$ mode pairs in the square 3C-SiC membrane. Additionally, we will conduct ringdown measurements for each distinct mechanical mode. The PSD provides the total decay rate, while the ringdown measurements yield the energy decay rate. By comparing the decay rates obtained from these two measurement techniques, we will demonstrate the exceptional performance of the square 3C-SiC membrane as a high-$Q$ mechanical resonator,particularly emphasizing its extremely low pure dephasing at low temperatures.

To leverage its outstanding frequency stability, we will employ a continuous pump-probe scheme, as shown in Fig.~1\textbf{d}, to demonstrate a new record in the passive storage of classical microwave signals. Furthermore, the pulsed pump-probe sequences depicted in Fig.~1\textbf{e} will enable on-demand ground-state cooling, writing, storage, and reading of microwave coherent states based on the long-lived phononic modes of the 3C-SiC membrane. Finally, we will also demonstrate coherent energy transfer between distinct mechanical modes.
\bigskip
\\
\noindent\textbf{Degeneracy-breaking and multiple high-$Q$ mechanical mode-pairs of a square 3C-SiC membrane}
\\
The device is mounted on a suspended oxygen-free copper plate, which is connected to the mixing chamber of a dry dilution refrigerator (with a minimum operating temperature of 8.9 mK) via copper braids. Without a mechanical isolation system, the mechanical modes cannot be thermalized below 500 mK. We provide a detailed discussion of the mechanical isolation techniques in Supplementary Note 2. The sheet resistance of the grown 3C-SiC, measured at room temperature, is approximately 1 $\Omega \cdot \text{cm}$. It is essential to verify the compatibility of this low-resistance 3C-SiC semiconductor material with microwave radio frequency circuits, particularly regarding whether this material will introduce significant dissipation to the superconducting cavity mode. As shown in Fig.~1\textbf{c}, the probe field can be used to detect the cavity responses under different external pump powers. In the low pump power regime (e.g., $P=-30$~dBm), the measured intrinsic and external decay rates are $\kappa_{\textrm{in}}/2\pi$~=~80~kHz and $\kappa_{\textrm{ex}}/2\pi$~=~120.25~kHz, respectively. Under strong pump power (e.g., $P=20$~dBm), the cavity total linewidth decreases to $\kappa/2\pi=160$~kHz. The frequency of the resonant cavity is quite stable. As the number of photons in the cavity increases, the resonant frequency changes only around 3~kHz. The cavity $S_{21}$ measurements indicate that our 3C-SiC membrane is well-compatible with superconducting resonators for microwave applications.
\begin{figure*}[ptb]
\centering
\includegraphics[scale=0.95]{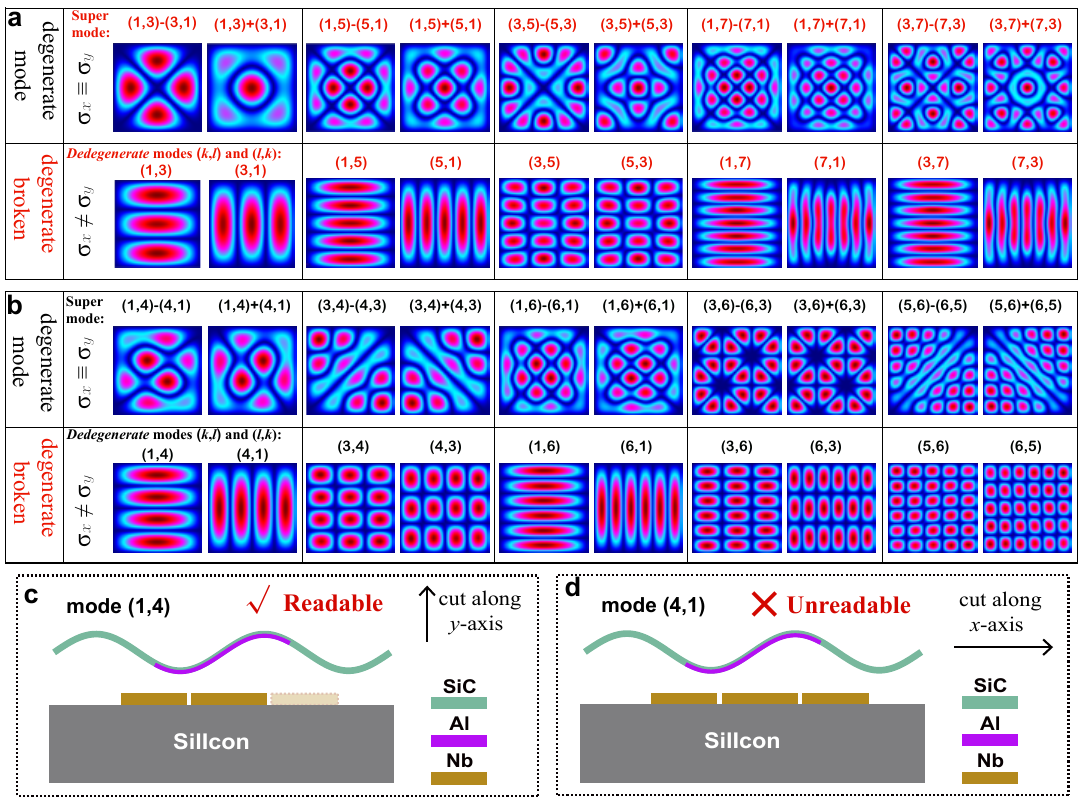}
\caption{\noindent\textbf{The mode shapes and effective mechanical capacitance for a square membrane in the stress governed regime.} The first (second) row in~\textbf{a} shows the ($k$,$l$) mode shapes with uniform (nonuniform) biaxial tensile stress $\sigma_{x}\equiv\sigma_{y}$ ($\sigma_{x}\neq\sigma_{y}$). The mode number $k$ and $l$ presented in ~\textbf{a} are both odd number. ~\textbf{b} presents the mode shapes for mechanical ($k$,$l$) modes with odd mode number $k$ and even number $l$. Similarly to ~\textbf{a}, the first and second rows correspond to uniform and nonuniform biaxial tensile stress cases, respectively. For odd \( k \) and even \( l \), \textbf{c} and \textbf{d} schematically illustrate how the vertical mechanical oscillations of the degenerate-broken ($k$,$l$) and ($l$,$k$) modes induce changes in the corresponding mechanical capacitance, using \( k = 1 \) and \( l = 4 \) as an example.}
\label{fig3}
\end{figure*}

We now read out these mechanical modes through electromechanical interactions by applying a weak continuous microwave pump tone to the cavity field via the SMA-type (depicted in Fig.~1\textbf{a}) input port of the 3D microwave resonator. The driving frequency is red-detuned to the cavity resonance by each mechanical resonance frequency, i.e., $\Delta=-\omega_{\textrm{m}}^{i}$. The beam-splitter-like interaction terms $H_{\textrm{int}}^{i}/\hbar=G_{i}(a^{\dag}b_{i}+b_{i}^{\dag}a)$ indicate that the voltage signal of output microwave field contains information such as the mechanical resonance frequency and cavity-field dressed mechanical linewidth.

Fig.~2\textbf{a} shows the measured power spectrum density (PSD) of the output microwave field in a wide frequency range (up to 2.37~MHz). For this device, 21 mechanical modes coupled to the cavity field have been measured. The $Q$-factors for each mechanical mode are measured through stroboscopic ringdown technology (see Supplementary Note 3) and are presented in Fig.~2\textbf{b}. Remarkably, 19 of the 21 detected mechanical modes have quality factors exceeding $10^8$.

As shown in Fig.~1\textbf{b} and Fig.~1\textbf{c}, our studied electromechanical device is particularly sensitive to those mechanical modes that exhibit a central vertical displacement. For a square membrane, when the mode numbers \( k \) and \( j \) are equal and both are odd, the mode shapes of these special modes consistently display a central vertical displacement, making them relatively easier to read out electromechanically. The finite element method (FEM) solutions in COMSOL for the mode shapes of the detected modes, such as (1,1), (3,3), and (5,5), are presented around each measured mode peak in Fig.~2\textbf{a}. The fundamental drumhead (1,1) mode has a resonant frequency of $\omega_{\textrm{m}}^{(1,1)}/2\pi$=382.15~kHz and a $Q$-factor of $Q^{(1,1)}=5.76\times10^{7}$. The modes (3,3) and (5,5) have frequencies and \( Q \)-factors of \( \omega_{\text{m}}^{(3,3)}/2\pi = 1.17 \, \text{MHz} \) with \( Q^{(3,3)} = 2 \times 10^8 \), and \( \omega_{\text{m}}^{(5,5)}/2\pi = 1.96 \, \text{MHz} \) with \( Q^{(5,5)} = 3.5 \times 10^8 \), respectively.

The fundamental mode is followed by a pair of nearly resonant mechanical modes with mode numbers (1,3) and (3,1), having quality factors of \(1.4 \times 10^8\) and \(1.8 \times 10^8\), respectively. Additionally, we identify another set of four nearly resonant mode pairs with higher mode numbers: (1,5) and (5,1), (3,5) and (5,3), (1,7) and (7,1), as well as (3,7) and (7,3). Other detected mechanical modes of the square 3C-SiC membrane include (1,4), (3,4), (1,6), (5,4), (3,6), (5,6), and (1,8). In conclusion, these electromechanically readable distinct modes or mode pairs share a common characteristic: the mode number \(k\) must be an odd number.

Next, we focus on the physical mechanism of electromechanical reading of the near-resonant mode pairs in this square 3C-SiC membrane, specifically the pairs \((k,l)\) and \((l,k)\) as demonstrated in Fig. 2\textbf{a}, where both \(k\) and \(l\) are unequal odd numbers. Considering a square membrane with uniform biaxial tensile stress, e.g., widely studied amorphous silicon-nitride (SiN)~\cite{yuan2015silicon,Guria2024Resolving,Peterson2016Laser,Piergentili2021Absolute,Yang2020Phonon,Sementilli2022Nanomechanical}, the natural vibration frequencies of a square film in the stress-governed regime (membrane) in vacuum are given by
\begin{equation}
\omega_{\textrm{m}}^{(k,l)}/2\pi=\frac{1}{2L}\sqrt{\frac{\sigma}{\rho}(k^2+l^2)}, \label{fmembrane}
\end{equation}
with the result that $\omega_{\textrm{m}}^{(k,l)}=\omega_{\textrm{m}}^{(l,k)}$. Here, $\sigma$ is the biaxial tensile stress and $\rho$ is the density of the material. Equation~(\ref{fmembrane}) indicates that when \( k \neq l \), the modes \((k,l)\) and \((l,k)\) exhibit two different mode shapes, \(\Psi_{(k,l)}\) and \(\Psi_{(l,k)}\), but have exactly the same frequency, which is referred to as degenerate modes. Since the frequencies are indistinguishable, if the square membrane is excited at this frequency, both of these degenerate modes will oscillate simultaneously. The overall vibration response of the square membrane will be a combination of all the degenerate mode shapes, represented as super modes: $\Psi_{kl}^{\pm}=a\Psi_{(k,l)}\pm b\Psi_{(l,k)}$, where \(a^2 + b^2 = 1\). Clearly, the formed super modes share the same frequency as the original degenerate modes and remain degenerate.

The finite element method (FEM) simulations for the super modes \(\Psi_{kl}^{\pm}\) are illustrated in the first row of Figure 3\textbf{a}. Additionally, the mode shapes of the original degenerate modes \(\Psi_{(k,l)}\) and \(\Psi_{(l,k)}\) resemble the patterns shown in the second row of Figure 3\textbf{a}. When the modes \((k,l)\) and \((l,k)\) are degenerate, their mode shapes will be symmetric, rotated by \(90^{\circ}\). Interestingly, both these mode shapes feature a central vertical up-and-down vibrational displacement. The superposition of mode shapes $\Psi_{(k,l)}$ and $\Psi_{(l,k)}$ can result in two situations: constructive interference [$\Psi_{kl}^{+}=(\Psi_{(k,l)}+ \Psi_{(l,k)})/\sqrt{2}$] and destructive interference [$\Psi_{kl}^{-}=(\Psi_{(k,l)}-\Psi_{(l,k)})/\sqrt{2}$] of the vibration displacements. Constructive interference results in an enhancement of the central vertical displacement, whereas destructive interference leads to a reduction, ultimately eliminating the central vertical displacement. In conclusion, for unequal odd $k$ and $l$, only the constructive superposition $\Psi_{kl}^{+}$ retains a central vertical displacement, which can modulate the mechanical capacitance and be read out through microwave cavity electromechanical interaction in this work. It is important to emphasize that the above conclusion applies specifically under the condition of uniform tensile stress on a square membrane, where \(\sigma_{x} = \sigma_{y}\).

However, a significant difference between the above theoretical analysis and our experiment is that, we detect near-resonance mode-pairs (indicated by red-colored mode numbers in Fig.~2\textbf{a}) instead of a single super mode $\Psi_{kl}^{+}$ resulting from constructive superposition. For a square membrane resonator, when the tensile stress becomes non-uniform (i.e., $\sigma_{x}\neq\sigma_{y}$), the super modes \(\Psi_{kl}^{\pm}\) will be broken and reduce to near-resonance \((k,l)\) and \((l,k)\) mode-pairs with distinct mode shapes $\Psi_{(k,l)}$ and $\Psi_{(l,k)}$, respectively. In later sections, this phenomenon is referred to as ``\textit{dedegenerate}". The second row of Fig.~3\textbf{a} shows the FEM simulations of the mode shapes for the dedegenerate \((k,l)\) and \((l,k)\) mode pairs with \(\sigma_{x} \neq \sigma_{y}\). Remarkably, both dedegenerate mode pairs are characterized by central vertical displacement, which can induce changes in mechanical capacitance. Thus, the degeneracy-breaking process allows the cavity to electromechanically read out both mechanical modes within the near-resonance mode pair. This represents a significant difference compared to a membrane with uniform biaxial stress, where only the constructive supermodes can be electromechanically read out.

It is worth noting that even if the stress distribution of a thin membrane, such as a well-studied SiN membrane~\cite{PhysRevLett.122.154301,Zhouxin2021HighQ,PhysRevLett.127.014304,PhysRevLett.128.153901}, is uniform at room temperature, stress redistribution and non-uniformity may occur during the cooling process of the refrigerator due to chip deformation. However, the stress distortion caused by this deformation often significantly reduces the quality factor of the mechanical modes. In our experiment, all detected mode pairs exhibit high-$Q$ behavior, with a quality factor \(Q\) exceeding \(10^{8}\). To further rule out this possibility, Fig.~2\textbf{c} and~\textbf{d} display the measured mechanical resonance and the corresponding $Q$-factors at room temperature, obtained using a laser Doppler vibrometer with the laser spot focused near the center of the square membrane. The measurement setup and laser PSD result of each detected mechanical modes are presented in Supplementary Note 4. Due to the fact that the laser spot cannot be exactly located at the center of the square membrane, the method using a Doppler vibrometer can detect more mechanical modes than the method using microwave cavity optomechanical interaction. We can still observe near-resonance mechanical modes at room temperature that correspond to the results detected at 10 mK, as presented in Fig.~2\textbf{a} and \textbf{b}. It is evident that the mechanical modes \((k, l)\) and \((l, k)\) are originally non-degenerate, and that the 3C-SiC membrane initially exhibits non-uniform biaxial tensile stress from room temperature. Additionally, Figure 2 illustrates that the \(Q\) values of all mechanical modes improve by three orders of magnitude as the environmental temperature decreases from room temperature to approximately 10 mK.

The biaxial tensile stresses used in the COMSOL simulation shown in Figure 3 are $\sigma_{x}=242$~MPa, and $\sigma_{y}=240$~MPa, respectively. The simulated frequencies show good agreement with the frequencies measured by the laser Doppler vibrometer. The finite element analysis also indicates that the dedegenerate initiates when difference in biaxial tensile stresses exceeds 0.1~Mpa. Once the stress difference reaches 1~Mpa, the dedegenerate mechanical modes ($k$,$l$) and ($l$,$k$) exhibits completely distinct mode shapes. In the special case where the odd mode number \( k \) equals \( l \), it is evident that, despite the biaxial tensile stress not being uniform, only a single mode exists, with no degenerate modes.
\begin{figure*}[ptb]
\centering
\includegraphics[scale=0.95]{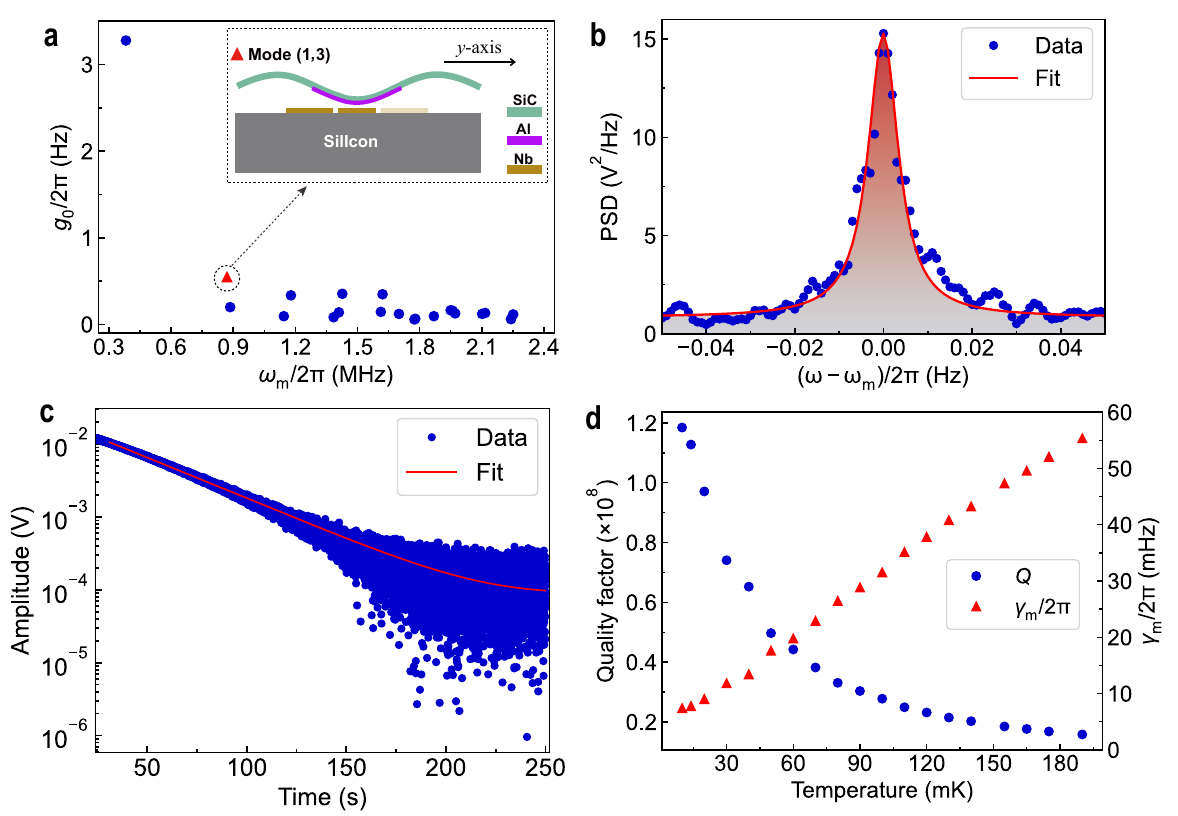}
\caption{\noindent\textbf{Electromechanical coupling strengthes and $Q$-values as a function of refrigerator temperature sweep.}~\textbf{a} The single-photon coupling strength between each mechanical mode to the common microwave cavity field. The inset in \textbf{a} illustrates how the vertical displacement of mode (1,3) affects the mechanical vacuum-gap capacitance, with a cut along the \textit{y}-axis.~\textbf{b} The PSD of the mechanical mode (1,3) obtained from the IQ measurement with a 1000~seconds sampling time.~\textbf{c} The mechanical energy ring-down shows a decay rate of $\gamma_{\textrm{m}}/2\pi=8.13$~mHz. The measured mechanical resonant frequency is $\omega_{\textrm{m}}/2\pi=875.318$~kHz and the spectrum linewidth via fit is obtained as $\gamma_{\textrm{m}}/2\pi=8.2$~mHz, corresponding to a quality-factor of $Q=1.18\times10^{8}$.~\textbf{d} The linewidth and $Q$-value of the mechanical mode with the strongest coupling to the microwave cavity (marked by blue-dot in ~\textbf{a}) evolve with the refrigerator's operating temperature.}
\label{fig4}
\end{figure*}

As discussed in previous sections, the central area of 3C-SiC square membrane is metallized (as shown in Fig.~1\textbf{a}), making the change in mechanical capacitance particularly sensitive to mechanical modes with central drumhead shapes. However, we emphasize that the notch structure along the $y$-axis in the bottom electrode (see inset of Fig.~2a for the detail) of the mechanical capacitor enables the detection of higher-order mechanical modes, particularly those exhibiting significant vertical displacement around the notch area. We thus move to demonstrate how the degenerance-breaking affect the cavity electromechanical readout of the mechanical modes with one of the mode number $k$ or $l$ is even and the other is odd. In Fig.~3\textbf{b}, the first row illustrates the mode shapes under uniform tensile stresses, where the mode shape of degenerate-superposition generated super modes $\Psi_{kl}^{\pm}$ exhibit a $90^{\circ}$ rotational symmetry. Notably, although the bottom antenna electrode contains a notch structure, the simultaneous presence of peaks and dips around the central metallized region of the square 3C-SiC membrane cancels each other out, resulting in nearly unchanged mechanical capacitances. Therefore, when one of the mode numbers $k$ or $l$ is even and the other is odd, all these degenerate supermodes cannot be read out electromechanically through the cavity field.

Remarkably, the breaking of rotational symmetry combining with the asymmetric notch design of the capacitor bottom electrode, enables the selective readout from mechanical modes ($k$,$l$) and ($l$,$k$). As shown in Fig.~2\textbf{a}, dedegenerate modes include (1,4), (3,4), (1,6), (5,4), (3,6), (5,6), and (1,8) are read out with $Q$-factors over $10^{8}$. For odd $k$ and even $l$, the FEM simulation and the mode shapes for mode pair ($k$,$l$) and ($l$,$k$) under non-uniform biaxial tensile stress are shown in the second row of Fig.~3\textbf{b}. Owing to the notch that extends upward along the \(y\)-axis of the bottom electrode, only the \((k, l)\) modes from the dedegenerate mode pair, defined by odd \(k\) and even \(l\) mode number, exhibit significant variations in mechanical capacitance, thereby facilitating their detection by the microwave cavity. Here the positive integer $l$ should be greater than 2 because, for the (1,2) mode, the area with the largest amplitude is not covered by the metallized aluminum electrode. In contrast, the \((l, k)\) modes do not induce significant capacitance changes and therefore cannot be detected.

Using \( k = 1 \) and \( l = 4 \) as an example, for mode (1,4), the standing wave nodes are distributed along the \( y \) direction, which is why the schematic diagram in Fig.~3\textbf{c} employs a cross-sectional view along the \( y \) axis. The mode vertical displacement combing the bottom antenna notch introduce the mechanical capacitance changes. In contrast, for the \((4, 1)\) mode, the standing wave nodes are distributed along the \(x\) direction, necessitating a cross-sectional view along the \(x\) axis in the schematic diagram as shown in Fig.~3\textbf{d}. However, the capacitance changes resulting from the vertical displacement cancel each other out, rendering the \((4, 1)\) mode undetectable by the microwave cavity. In the final scenario, when both mode numbers are even, despite the occurrence of degeneracy breaking, the detectability of these modes is constrained by negligible capacitance changes, rendering them undetectable.
\bigskip
\\
\noindent\textbf{Multiple degeneracy-broken mechanical modes with extremely low pure dephasing}
\\
The strength of optomechanical interactions in a microwave cavity electromechanical system can be quantified by the vacuum coupling rate $g_{0}$. For the detected 21 mechanical modes, the vacuum coupling rate for each mode are presented in Fig.~4\textbf{a}. Finite element modeling and a comparison between the simulation of $g_{0}$ and the measurement results are presented in Supplementary Note 5.

Although the fundamental (1,1) mode exhibits the strongest $g_{0}$, its $Q$-factor is only $5.7\times10^{7}$, which is the minimum among all the detected modes. Considering both the mechanical $Q$ value and the resolved-sideband condition, our subsequent studies will focus on the mechanical mode (1,3) highlighted in red in Fig.~4\textbf{a}. Meanwhile, the inset in Fig.~4\textbf{a} shows how the vertical displacement of mode (1,3) changes the mechanical vacuum-gap capacitance, with a cut along the \textit{y}-axis. This mode has a resonance frequency of $\omega_{\textrm{m}}/2\pi=871.318$~kHz and a remarkable $Q$-factor of $1.18\times10^{8}$.

Using Eq.~(\ref{PSDdegrades}) to fit the measured PSD yields the mechanical decay rate for each mechanical mode. As an example, Fig.~4\textbf{b} shows the measured PSD spectrum of mechanical mode (1,3), and the fitted energy decay rate is of $\gamma_{\textrm{m}}/2\pi$~=~8.2~mHz. Simultaneously applying driving and probe tone with a frequency difference by $\omega_{\textrm{m}}$ can effectively excite the mechanical motion. After the driving tone is abruptly turned off, the ring-down trace of the fundamental drumhead mode is shown in Fig.~4\textbf{c}. Using exponential fitting, we obtain the energy decay of the mechanical resonator with a rate of $\gamma_{\textrm{m}}/2\pi=8.13$~mHz, which agrees well with the value obtained from the PSD measurement in Fig.~4\textbf{b}. In addition to the energy decay rate and lifetime, pure dephasing rate is also an important parameter to be considered.

Although the energy spectrum tests and the time-domain ringdown tests suggest that pure dephasing is a very small effect, we will record the ringdown data for an extended duration using I/Q sampling~\cite{yuan2015silicon}. By directly comparing the linewidth of the energy spectrum obtained after FFT transformation with the energy relaxation rate derived from direct fitting (see Supplementary Note 6), we can accurately determine the magnitude of pure dephasing, which is measured at $\gamma_{\varphi}=0.28$~mHz. These results indicate that dephasing does not significantly contribute to decoherence in the SiC square membrane resonators.

Recalling the measurements made by the Doppler vibrometer at room temperature, the $Q$ factor for mode (1,3) was only 2.6$\times10^{4}$. This performance underwent an impressive improvement of almost four orders of magnitude when operating at a low temperature around 10~mK. To investigate how the decrease in temperature affects the evolution of the $Q$-factors, a temperature sweep experiment was conducted. The measured $Q$-factors of mode (1,3) as a function of the dilution refrigerator operating temperature are depicted in Fig.~4\textbf{d}. The linewidth ($Q$-factor) of the mechanical oscillator continues to decrease (increases) as the operating temperature decreases, showing a linear dependence. For every 10~mK increase in temperature, the linewidth increases by 2.66~mHz. Furthermore, the frequency of the mechanical oscillator remains highly stable across the tested temperature range of 10 mK to 190 mK, exhibiting a variation of less than 1~mHz over a duration of one day.

For instance, the focused study on the mechanical mode \((1, 3)\) of the 3C-SiC square membrane reveals an energy decay time of of approximately $T_{1}$~=~19.9~seconds, a total linewidth of $\gamma_{\textrm{m}}/2\pi$~=~8.2~mHz, and particular with an extremely low pure dephasing rate of of $\gamma_{\varphi}$~=~0.28~mHz. Table 1 summarizes the pure dephasing of all detected mechanical modes and their proportion in the total linewidth. These initial characterizations enable us to conclude that the degeneracy-breaking 3C-SiC membrane supports multiple mechanical modes with exceptional performance, particularly characterized by extremely low pure dephasing. Motivated by this enhanced mechanical frequency stability, we proceed to demonstrate its potential application as a classical microwave memory device.
\begin{table*}[h]
    \centering
    \caption{The pure dephasing and its proportion of the total linewidth for each mechanical mode}
    \resizebox{0.65\textwidth}{!}{
    \begin{tabular}{ccccc} 
        \toprule
        mode number & frequency ($2\pi\cdot$) & total linewidth ($2\pi\cdot$) & pure-dephasing ($2\pi\cdot$) & proportion \\
        \midrule
        (1,1) & 382.147~kHz & 6.83~mHz & 0.22~mHz & 3.22\% \\
        (1,3) & 871.318~kHz & 8.2~mHz & 0.28~mHz & 3.41\% \\
        (3,1) & 885.413~kHz & 5.31~mHz & 0.4~mHz & 7.53\% \\
        (1,4) & 1.144~MHz & 7.9~mHz & 0.33~mHz & 4.17\% \\
        (3,3) & 1.178~MHz & 6.04~mHz & 0.25~mHz & 4.13\% \\
        (3,4) & 1.384~MHz & 7.55~mHz & 0.16~mHz & 2.11\% \\
        (1,5) & 1.41~MHz & 17.46~mHz & 0.27~mHz & 1.54\% \\
        (5,1) & 1.426~MHz & 8.56~mHz & 0.28~mHz & 3.27\% \\
        (3,5) & 1.612~MHz & 5.26~mHz & 0.28~mHz & 5.32\% \\
        (5,3) & 1.621~MHz & 7.1~mHz & 0.23~mHz & 3.23\% \\
        (1,6) & 1.7~MHz & 10.05~mHz & 0.22~mHz & 2.18\% \\
        (5,4) & 1.776~MHz & 8.09~mHz & 0.37~mHz & 4.57\% \\
        (3,6) & 1.868~MHz & 7.28~mHz & 0.23~mHz & 3.15\% \\
        (1,7) & 1.949~MHz & 7.75~mHz & 0.26~mHz & 3.35\% \\
        (5,5) & 1.959~MHz & 5.88~mHz & 0.32~mHz & 5.44\% \\
        (7,1) & 1.971~MHz & 9.7~mHz & 0.31~mHz & 3.19\% \\
        (3,7) & 2.101~MHz & 8.91~mHz & 0.31~mHz & 3.47\% \\
        (7,3) & 2.116~MHz & 8.38~mHz & 0.18~mHz & 2.14\% \\
        (5,6) & 2.24~MHz & 11.72~mHz & 0.5~mHz & 4.26\% \\
        (1,8) & 2.25~MHz & 22.58~mHz & 0.62~mHz & 2.74\% \\
        \bottomrule
    \end{tabular}
    }
\end{table*}
\bigskip
\\
\noindent\textbf{Hour-level group-delay and slow-light with extremely stable mechanical modes}\\
The pure dephasing of the studied 3C-SiC square membrane is measured at only 0.28 millihertz, constituting just 3.5\% of the total linewidth. In comparison, the pure dephasing value for silicon nitride (SiN), which has made significant advancements in improving quality factors, accounts for over 25\% of the total linewidth~\cite{yuan2015large,liu2023coherent,fink2016quantum,seis2022ground,yuan2015silicon}. This results in at least an 86\% reduction in frequency instability for the 3C-SiC resonator compared to SiN resonators. Consequently, our 3C-SiC resonator exhibits remarkable frequency stability, as demonstrated by a frequency shift of less than 1 millihertz per day. The narrow linewidth of SiN mechanical resonators have resulted in a reported maximum group delay on the order of hundreds of seconds~\cite{liu2021optomechanical}. In this context, the limitation for high-\textit{Q} mechanical resonators is often frequency instability. In the following section, we will demonstrate how the significantly improved frequency stability of the focused 3C-SiC square membrane in this work can contribute to setting a new record in the passive storage of propagating microwave signals.
\begin{figure*}[ptb]
\includegraphics[scale=0.81]{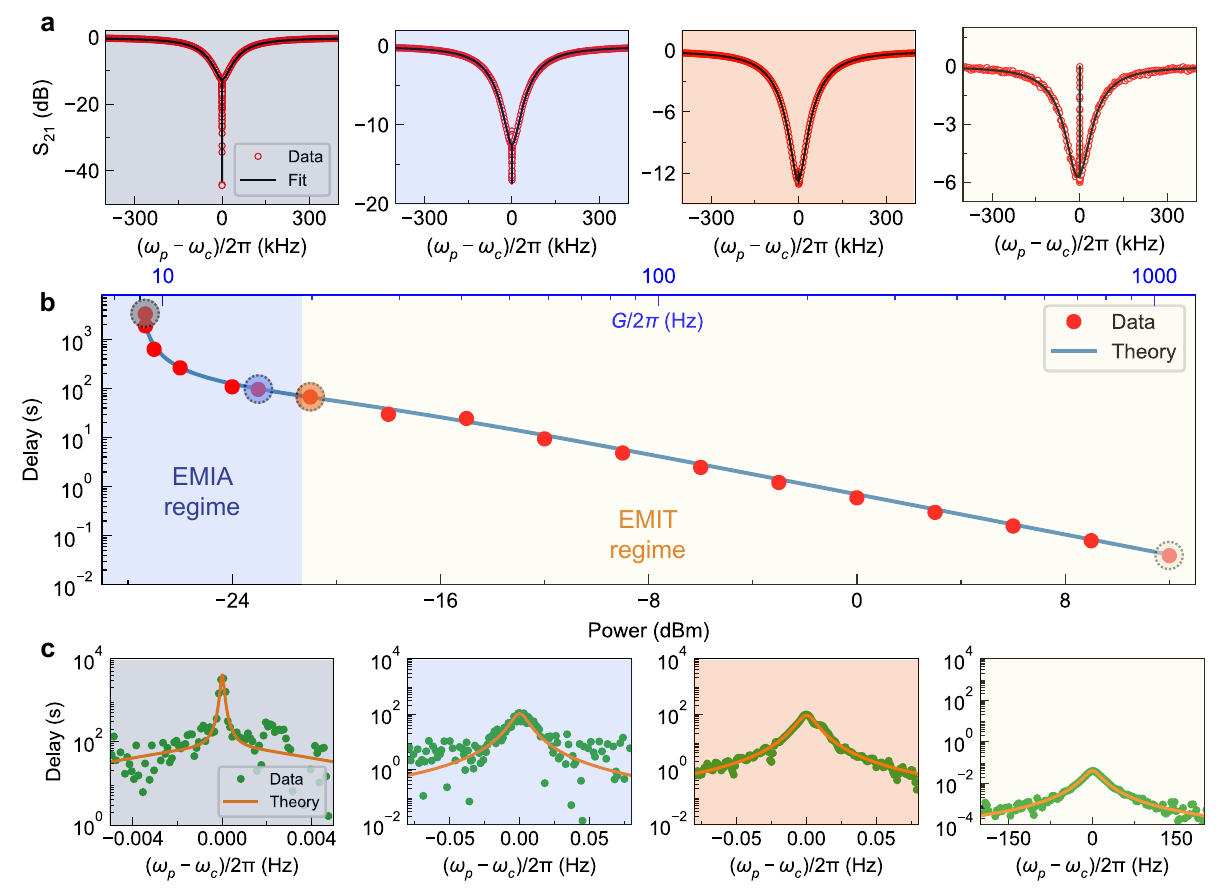}
\caption{\noindent\textbf{Continuous-wave Pump-probe experiment and ultralong group-delay and slow-light.} \textbf{a} The $S_{21}$ transmission spectrum as a function of probe-field frequency detuning $\delta=\omega_{\textrm{p}}-\omega_{\textrm{c}}$. Red circles are experimental data and the black-solid curves are the fitting curves. \textbf{b} shows the delay time for a set of driving-tone powers. The red-solid circles represent the experimental data and the blue curve is the theory curve calculated from parameters fitting from \textbf{a}. Here, ``dBm'' is an arbitrary unit and refers to the generator power. The dashed circles which are colored according to the minor panels in \textbf{a}, and \textbf{c}, respectively. The use of blue and beige background colors in \textbf{b} signifies EMIA and EMIT regimes, respectively. \textbf{c} the delay time as a function of probe-tone frequency detuning $\delta$, wherein green dots and orange curves are experimental data and fitting curves, respectively. For \textbf{a} and \textbf{c}, from left to right the pump power is gradually increased and the corresponding slow-light time is marked by dashed circles, as shown in \textbf{b}.}
\label{fig5}
\end{figure*}

We now demonstrate the transmission response of a weak probe-tone with frequency $\omega_{\textrm{p}}$ and amplitude $\varepsilon$, also well known as the continuous field pump-probe experiment. The scheme has been depicted in Fig.~1\textbf{d}, where the pump-tone is red-detuned to cavity resonance frequency by $\Delta=-\omega_{\textrm{m}}$. The weak probe-tone is scanned by continuously changing frequency, and the dynamics of interest occur when the probe frequency enters in the microwave cavity resonance regime. Working in the rotating frame at the pump-tone frequency $\omega_{\textrm{d}}$, and considering the input-output relations, the transmission coefficient is given as
\begin{equation}
t=1+\frac{\eta \kappa \left(i\delta -\gamma_{\textrm{m}}/2\right)}{\left(i\delta-\gamma_{\textrm{m}}/2\right) \left(i\delta -\kappa/2\right)+G^{2}}. \label{Transimission}
\end{equation}
The amplitude and phase responses can be calculated with the relations $T=|t|^2$, and $\varphi=\arg \left(t\right)$, respectively. The group delay (slow-light time $\tau$) is obtained from the slope of phase-response, i.e., $\tau = \partial \varphi /\partial \omega_{\textrm{p}} = \partial \varphi /\partial \delta$. The transmission coefficient at zero-detuning ($\delta=0$) is given as
\begin{equation}\label{zerodetuning}
 T_{z}=\left|\frac{4G^2-(2\eta-1)\kappa\gamma_{\textrm{m}}}{4G^2+\kappa\gamma_{\textrm{m}}}\right|^{2},
\end{equation}
from which we find that the prerequisite for coherent perfect absorption (i.e., $T_{z}=0$) is $\eta>1/2$, viz., external overcoupling $\kappa_{\textrm{ex}}>\kappa_{\textrm{in}}$. For our device, the external coupling coefficient is determined to be $\eta=0.6$. Fig.~5\textbf{a} shows the probe-tone $S_{21}$ transmission response under a set of electromechanical couplings strengths $G/2\pi=[10.46,15.85,19.2,961.07]$~Hz, respectively. Under a weak coupling strength (e.g., $G/2\pi=10.46$~Hz), the probe-tone transmission spectrum initially shows an additional absorption dip which is more than 30~dB lower than cavity $S_{21}$ background base-line. With coupling rate $G$ increasing, the absorption dip starts to rise until it reaches the background of the microwave cavity $S_{21}$ curve, and finally, a transparency window appears.

The observed electromechanical-induced-absorption (EMIA) and -transparency (EMIT) phenomena can be understood as a result of constructive and destructive interference between two different pathway transitions~\cite{liu2021optomechanical,agarwal2010electromagnetically,weis2010optomechanically,safavi2011electromagnetically,karuza2013optomechanically,dong2015brillouin,kim2015non,zhou2013slowing,fan2015cascaded,hocke2012electromechanically,xiong2018fundamentals,liu2017electromagnetically}. When probe-tone frequency arrives at cavity resonance, a beating (at a mechanical resonance frequency) between pump- and probe-tones coherently excites the mechanical oscillations. Based on the electromechanical coupling, the pump-tone creates a Stokes sideband by absorbing the mechanical excitations. Concurrently, the accompanying anti-Stokes process is substantially suppressed as it is energetically situated below the cutoff frequency of microwave cavity. Additionally, the amplitude of this Stokes sideband depends on the coupling strength, which can be used to control the evolution from EMIA to EMIT.

We now focus on the interference between the photons of the Stokes sideband and probe-tone inside the microwave cavity. When electromechanical coupling rate equals the critical coupling strength $G_{\textrm{c}}=\sqrt{(2\eta-1)\kappa\gamma_{\textrm{m}}/4}$, all created Stokes sideband photons constructively interfere with input probe-tone photons inside the microwave cavity (corresponding to destructive interference outside the cavity), leading to a coherent perfect absorption of transmitted probe-tone (e.g., as shown in the first row of Fig.~5\textbf{a}). Above critical coupling ($G >G_{c}$), the number of converted Stokes-sideband photons now is more than the number of probe-tone photons. Therefore, only part of the converted Stokes-sideband photons constructively interfered with the probe-tone photons (absorbing them). The remaining Stokes-sideband photons are recorded and the bottom of the absorption dip starts to rise (e.g., as shown in the second and third row of Fig.~5\textbf{a} and \textbf{b}). While further increasing the electromechanical coupling, the converted Stokes-sideband photons are strong enough to increase the transmission beyond the bare microwave cavity background, as shown in the fourth row of Fig.~5\textbf{a}. As a result, an electromechanically-induced transmission window is observed.
\begin{figure*}[pth]
\includegraphics[scale=0.96]{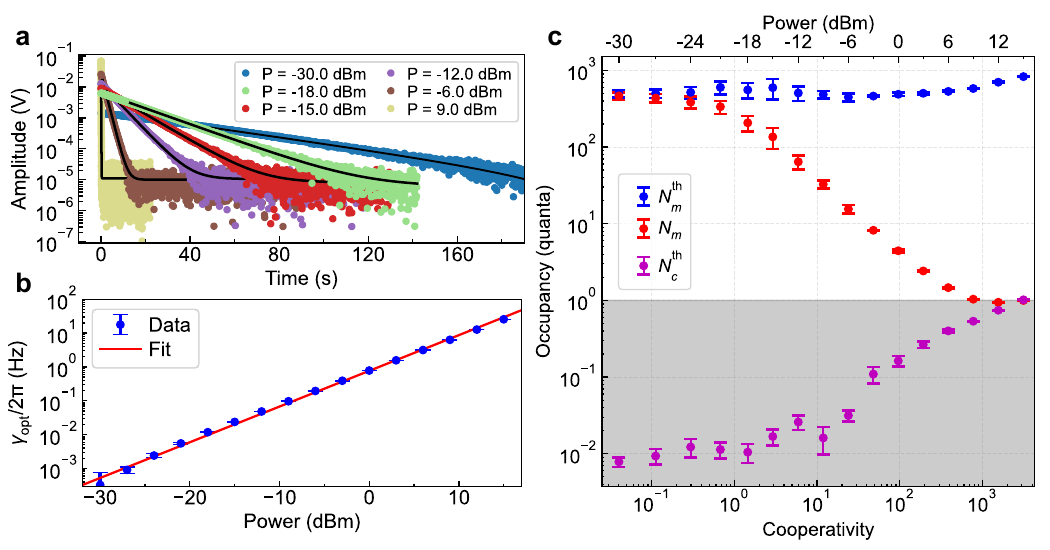}
\caption{\noindent\textbf{Sideband cooling of the mechanical memory mode.} \textbf{a} The ringdown measurement of total mechanical linewidth $\gamma_{\textrm{tot}}$ with a set of sideband cooling powers $P$. Colored-solid dots are experiment data and black-solid curves represent the fitting. \textbf{b} Cavity-field backaction introduces an additional optical-damping $\gamma_{\textrm{opt}}$ for the mechanical resonator. $\gamma_{\textrm{opt}}$ exhibits a linear growth with increasing the sideband driving power $P$. \textbf{c} The mechanical occupation is calibrated as the number of motional quanta. $N_{\textrm{m}}^{\textrm{th}}$ ($N_{\textrm{c}}^\textrm{th}$) is the effective thermal bath occupation of the mechanical mode (microwave cavity mode), respectively. Error bars equal one standard deviation.}
\label{fig6}
\end{figure*}

The measured group delays (slow-light time $\tau$) as a function of electromechanical coupling rate are shown in Fig.~5\textbf{b}. With increasing the electromechanical coupling rate $G$, group delay shows a decreasing trend. We noticed that the rate of group delay decrement changes dramatically when the coupling strength $G$ is in the range where EMIA occurs. However, when $G$ enters the range where EMIT occurs, the rate of group delay decrement remains unchanged. Under the condition of critical coupling $G=G_{\textrm{c}}$, coherent perfect absorption occurs ($T_{z}=0$), and at the same time, a singular point appears in the group delay with infinity slow-light or fast time~\cite{liu2021optomechanical}. Across the critical coupling strength \(G_{\textrm{c}}\), there is an abrupt transition from infinite group advance to delay. This transition leads to a dramatic change in group delay as the coupling strength \(G\) bypasses the singularity. In experimental scenarios, mechanical frequency jitter causes the critical coupling strength condition for achieving infinite group delay to vary, presenting significant challenges. Conversely, the maximum group delay that can be measured in experiments is limited by the frequency stability of the mechanical modes. Here, we report a new record for group delay time, attributed to the inherent advantages in frequency stability. As shown in Fig.~5\textbf{b}, the longest delay time we measured in the experiment is 4035~seconds (more than an hour), achieved through the high \textit{Q}-factors and extremely low pure dephasing rate of 3C-SiC square membrane. Leveraging the exceptional mechanical frequency stability, this interface enables tunable light slowing, achieving group delays of up to an impressive \emph{hour}. This significant improvement over previous pioneering works marks a substantial advancement in optomechanical group delay and slow light technologies~\cite{kristensen2024long,fink2016quantum,teufel2011circuit,zhou2013slowing,lake2021processing,shandilya2021optomechanical,merklein2017chip,JingHui2023Magnon,YingWu2019Bose}. Fig.~5\textbf{c} shows the measured group delay versus probe-tone detuning $\delta$. The maximum delay time occurs at the zero-detuning point and delay-time gradually decreases as the coupling rate increases. When the normalized transmission coefficient is close to unity, the energy transfer efficiency reaches its maximum. At the highest transmission efficiency, we emphasize that the delay time can still reach 40.9~ms, which is the longest delay time achieved~\cite{agarwal2010electromagnetically,weis2010optomechanically,safavi2011electromagnetically,karuza2013optomechanically,dong2015brillouin,kim2015non,zhou2013slowing,fan2015cascaded,hocke2012electromechanically,xiong2018fundamentals,liu2017electromagnetically,liu2021optomechanical}, to the best of our knowledge.
\bigskip
\\
\noindent\textbf{Sideband cooling of the \textit{dedegeneracy} mechanical modes}\\
The above continuous-fields-based pump-probe experiments mainly demonstrate the performance of our device in storing microwave signals for a long time. For classical applications, the thermal noise in cavity field or mechanical mode can be neglected. However, for quantum storage, one major obstacle is that the thermal phonons add noise into coherent mechanical excitations during the storage processes. We now show effective optomechanical sideband cooling to prepare mechanical oscillators toward their quantum ground state.

We add two layers of a mechanical damper to isolate external mechanical vibration noises (see Supplementary Note 2), including the pulse tube vibrations. We assume that the phonon occupation of this fundamental mechanical mode can arrive at its equilibration with environment temperature at several hundred millikelvins (mK) provided by the cryostat. Using the Bose-Einstein statistics for mechanical thermal states, i.e., $N_{\textrm{m}}^{\textrm{th}}=(e^{\hbar\omega_{\textrm{m}}/k_{\textrm{B}}T}-1)^{-1}$, we can extract a calibration constant between the area of mechanical sideband spectrum and the thermal bath phonon occupations in quanta. Here, $k_{\textrm{B}}$ is the Boltzmann constant constant and $T$ is the working temperature of the cryostat. We find that the peak area grows linearly with $T$ (see Supplementary Note~7) when the refrigerator's working temperature is above 20~mK, corresponding to an initial phonon occupation of $N_{\textrm{m}}^{\textrm{th}}=476$.

To cool the mechanical mode, we continue to use the red-sideband drive with an optimal detuning of $\Delta=-\omega_{\textrm{m}}$. The scheme can be depicted as that demonstrated in Fig.~1\textbf{d}, but without the probe-tone. In the resolved-sideband regime, the PSD of the output cavity field is finally expressed as the formula given in Eq.~(\ref{PSDoutputfinal}). Fig.~6\textbf{a} shows mechanical ringdown curves under different sideband cooling power $P$. In the experiment, the ringdown lifetime starts to decrease as cooling power $P$ increases. This can be intuitively understood as the additional optical-damping broadens the linewidth of the mechanical mode, i.e., $\gamma_{\textrm{tot}}=\gamma_{\textrm{opt}}+\gamma_{\textrm{m}}$. We can obtain the total mechanical damping rate $\gamma_{\textrm{tot}}$ by fitting the ring-down curves as shown in Fig.~6\textbf{a}. Fig.~6\textbf{b} shows that the extracted optical-damping $\gamma_{\textrm{opt}}$ depends linearly on increasing the sideband cooling power $P$, which is in good agreement with the theoretical predictions.
\begin{figure*}[pth]
\includegraphics[scale=0.97]{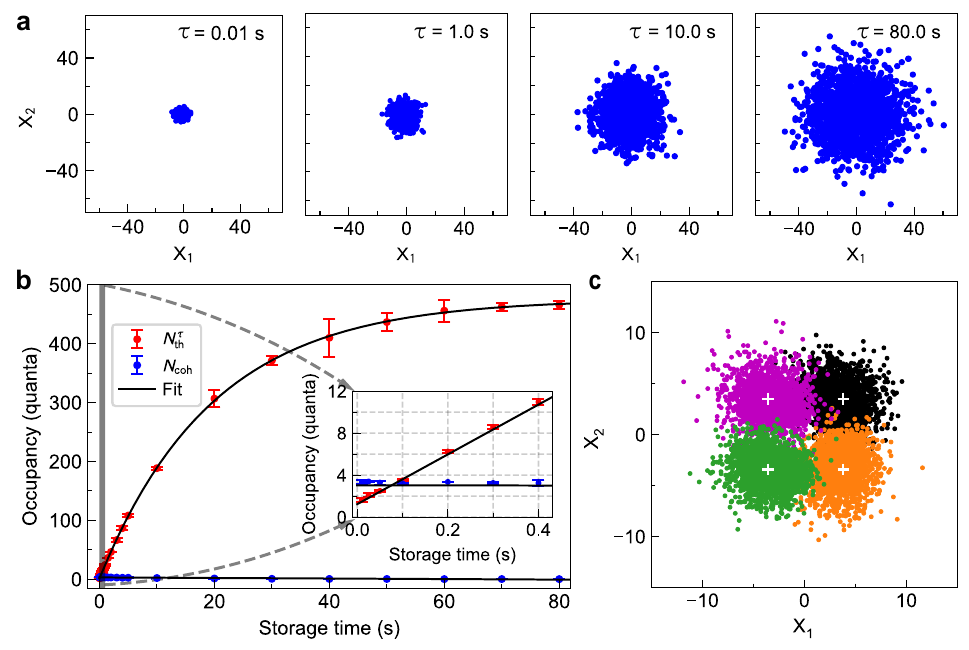}
\caption{\noindent\textbf{On-demand microwave state storage and tomography in the time domain.} \textbf{a} Scatter plots of the measured quadratures of motion with coherent input signals for different evolution times $\tau=0.01,~1,~10,~80$~s. One pulse-sequence measurement yields a point (blue dot) in the quadrature phase space. Each figure in \textbf{a} is obtained by repeating the pulse-sequence measurements 3000~times. The offset of the center of the scatter plot relative to the origin of the coordinates represents the component of the coherent excitations $N_{\textrm{coh}}$. \textbf{b} The coherent and thermal components of the retrieved microwave pulse after a storage time of $\tau$. The green triangles (red dots) are experimental data of thermal (coherent) occupation $N_{\textrm{th}}^{\tau}$ ($N_{\textrm{coh}}$) with error bars. At the same time, the solid lines are the corresponding exponential fit. \textbf{c} The phases of input signal pulses in the protocol of recovered microwave pulse with a phase increment of $\pi/2$.}
\label{fig7}
\end{figure*}

Ideally, as the sideband cooling power increases, the number of phonons gradually decreases and the oscillator enters the ground state. However, in the experiment, the cooling may be limited by microwave heating which raises the bath temperatures of the cavity and of the mechanical resonator. The thermal occupations $N_{\textrm{c}}^\textrm{th}$ and $N_{\textrm{m}}^\textrm{th}$ thus become dependent on the sideband pump power. Noise spectra of cavity emission and typical mechanical sidebands are presented in Supplementary Note 8. Using the function given in Eq.~(\ref{PSDoutputfinal}) and through the nonlinear least-squares fitting of the PSD, we can obtain the values of $N_{\textrm{c}}^\textrm{th}$ and $N_{\textrm{m}}^\textrm{th}$ under different sideband cooling power or optomechanical cooperativity. Using the solutions for $b[\omega]$, the phonon occupation $N_{\textrm{m}}$ can be derived via equipartition theorem~\cite{dobrindt2008parametric,rocheleau2010preparation}. Considering the case of optimal detuning $\Delta=-\omega_{\textrm{m}}$, and $\kappa^{2}\gg(4G^2,\kappa\gamma_{\textrm{m}},\gamma_{\textrm{m}})$, we find
\begin{equation}\label{finalphonon}
N_{\textrm{m}}=\frac{\gamma_{\textrm{m}}}{\gamma_{\textrm{opt}}+\gamma_{\textrm{m}}}N_{\textrm{m}}^\textrm{th}+\frac{\gamma_{\textrm{opt}}}{\gamma_{\textrm{opt}}+\gamma_{\textrm{m}}}N_{\textrm{c}}^{\textrm{th}}.
\end{equation}
The final steady-state occupations of mechanical mode ($N_{\textrm{m}}$), the thermal occupations of cavity field ($N_{\textrm{c}}^\textrm{th}$) and mechanical mode ($N_{\textrm{m}}^\textrm{th}$) bathes, as a function of both the cooling power $P$ and optomechanical cooperativity, are summarized in Fig.~6\textbf{c}.

Equation~(\ref{finalphonon}) shows that the sideband cooling can never reduce the mechanical mode occupancy below the occupancy of the cavity thermal bath. As shown in Fig.~6\textbf{c}, the value of $N_{\textrm{m}}$ depends mainly on the final occupancy number of cavity thermal bath $N_{\textrm{c}}^\textrm{th}$ at the highest powers. In this experiment, the minimum average phonon occupation is $N_{\textrm{m}}=0.9\pm0.03$. Towards achieving the ground-state cooling, cavity thermal occupancy $N_{\textrm{c}}^\textrm{th}$ must be controlled within one quantum. For currently reported cavity electromechanical systems, such as those based on either crystalline silicon acoustic cavities~\cite{maccabe2020nano} or amorphous SiN membrane oscillators~\cite{liu2022quantum,seis2022ground,higginbotham2018harnessing,liu2023coherent}, significant heating effects are observed in the mechanical thermal baths. However, for the 3C-SiC studied in this work, the fundamental drumhead mode of the membrane resonator does not exhibit a significant heating effect, which can be attributed to the extremely high thermal conductivity of SiC in 3C phase.
\bigskip
\\
\noindent\textbf{On-demand state capture, storage and retrieval}\\
We then investigate the device's performance as an on-demand phononic memory by measuring the capture, storage, and retrieval of itinerant microwave pulses in the time domain. Based on the beam-splitter type optomechanical interaction $H_{\textrm{BS}}=G\left(a^{\dag}b+b^{\dag}a\right)$, the state encoded in an itinerant microwave pulse can be written into and read out of the mechanical memory mode by controlling the pulse sequence of the transfer field. A constant sideband cooling tone is added to initialize the mechanical memory to its ground state. The corresponding optical-damping rate is $\gamma_{\textrm{opt}}^{\textrm{g}}=12.6$~Hz. The state capture, storage, and retrieval pulse-sequence has been depicted in Fig.~1\textbf{e}. It has been experimentally verified in the literature that matching signal growth rate $\Gamma$ with sideband cooling rate can result in an optimal state recover efficiency~\cite{palomaki2013coherent,liu2023coherent}. So, in this stage, we generate a microwave pulse characterized by an amplitude that exhibits exponentially growth at a specified rate of $\Gamma=\gamma_{\textrm{opt}}^{\textrm{g}}$.

The coherent state arrives at the cavity resonator at the moment of $t_{1}$. From $t_{1}$ to $t_{2}$ (duration 60~ms), when a coherent state arrives inside the microwave cavity, the constant sideband cooling tone begins to play another role, that is transferring the amplitude and phase information of such input signal-pulse into the phononic memory mode of 3C-SiC membrane resonator. As the microwave coherent state is captured, a coherent portion of the mechanical excitations continuously increases. In this stage, the sideband cooling tone is also called as the write field. At the moment of $t_{2}$, the writing field is closed, and the captured microwave state is stored as coherent phonons inside our memory device. After a certain period of storage ($\tau$), constant sideband cooling tone is added again (60~ms from $t_{3}$ to $t_{4}$), which maps the state from phononic mode back to an itinerant microwave pulse. The constant sideband cooling tone retrieves the microwave coherent state and thus is called the readout field in the recovering stage. The pulse sequence is completed after the time of $t_{4}$. During the entire process, the itinerant microwave coherent state has experienced the process of capture, storage, and retrieval. In the readout stage, we use an optomechanical phase-insensitive amplification technique (with a blue-detuned sideband pump) to improve the signal-to-noise ratio~\cite{massel2012multimode,Youssefi2023squeezed,reed2017faithful}.
\begin{figure*}[pth]
\includegraphics[scale=0.85]{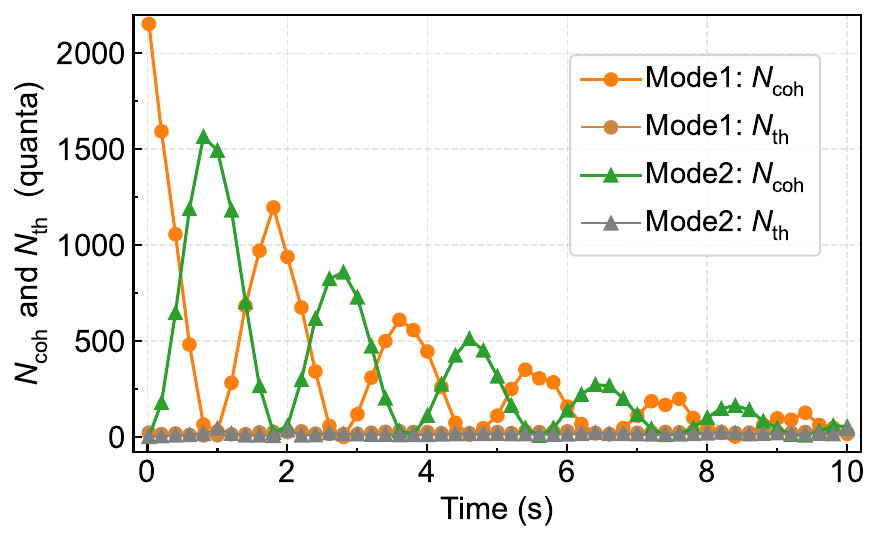}
\caption{\noindent\textbf{Optomechanical state swapping occurs between two mechanical modes, both of which are initially sideband cooled to their ground state.} The measurement of the state-swapping pulse sequence, repeated a thousand times, yields scattering points in quadrature phase space that enable us to calibrate the coherent and thermal components of the phonon numbers for each mechanical mode used in the state swapping.}
\label{fig8}
\end{figure*}

During the storage procedure, the phononic mode undergoes a rethermalization process. We now directly present the coherent storage lifetime by recording its thermal decoherence in the time domain. Turning on the readout field, the amplitude and phase of $V_{\textrm{out}}$ encode the state of the mechanical mode after a storage time of $\tau$. The voltage signal $V_{\textrm{out}}$ of recovered itinerant microwave pulses are recorded and analyzed by a spectrum analyzer. The quadrature amplitudes of mechanical motion are obtained by projecting $V_{\textrm{out}}$ onto the local oscillator with frequency $\omega_{\textrm{out}}$:
\begin{equation}\label{quadratures}
\begin{aligned}
 X_{1}=C\sum_{i}V_{\textrm{out}}\left(t_{i}\right)\cos\left(\omega_{\textrm{out}} t_{i}\right),\\
 X_{2}=C\sum_{i}V_{\textrm{out}}\left(t_{i}\right)\sin\left(\omega_{\textrm{out}} t_{i}\right).
\end{aligned}
\end{equation}
The completion of each pulse-sequence measurement generates a single point in the quadrature phase space. We can obtain the state tomography of the retrieved microwave state by repeating the pulse-sequence thousands of times. From quadrature distributions, we can distinguish the coherent component of mechanical excitations $N_{\textrm{coh}}=\langle X_{1}\rangle^{2}+\langle X_{2}\rangle^{2}$ and incoherent component of thermal occupancy $N_{\textrm{th}}^{\tau}=\langle\left(X_{1}-\left\langle X_{1}\right\rangle\right)^{2}\rangle+\langle\left(X_{2}-\left\langle X_{2}\right\rangle\right)^{2}\rangle-n_{\textrm{add}}$. Brackets here represent an ensemble average. Given sufficient storage time, we postulate that the mechanical oscillator fully reverts to its initial thermal state, achieving a state of thermal equilibrium with a phonon occupation number of $N_{\textrm{m}}^{\textrm{th}}=476$. We thus can calibrate the gain-factor $C$ in Eqs.~(\ref{quadratures}) and express $X_{1,2}$ in units of quanta. The calibration of the gain factor and the added noise are presented in Supplementary Note 9.

We now demonstrate microwave coherent state storage through the coherent electromechanical interface and the phononic memory mode. The scatter plots of the measured quadratures of the recovered microwave pulses are presented in Fig.~7\textbf{a}, for different storage time $\tau=0s,1s,10s,80s$, respectively. As the storage interval $\tau$ increases, the variance $\sigma^{2}=N_{\textrm{th}}^{\tau}$ starts to increase due to the re-thermalization of the phononic memory mode. Fig.~7\textbf{b} shows the measured coherent and thermal components of the mechanical memory mode, as a function of the storage time interval $\tau$. The mechanical memory mode takes more than 70~seconds for the free evolution from its ground state to a final thermal equilibrium state. The average photon number of the captured microwave coherent state is 4.2~quanta. The coherent component of phononic mode decays exponentially and the fit of $N_{\textrm{coh}}$ results in $\gamma_{\textrm{m}}/2\pi=8.12\pm0.19$~mHz, corresponding to an energy decay rate of $T_{1}=19.9\pm0.3$~seconds. These experiment data show excellent agreement with the predicted evolution. Coherent storage lifetime is an important figure of merit for characterizing a quantum memory. Zoomed-in area in Fig.~7\textbf{b} shows the thermalization and energy decay of the mechanical memory in shorter evolution intervals. The storage lifetime is determined by the thermal decoherence rate of the mechanical memory, which is calibrated as $\Gamma_{\textrm{th}}/2\pi=3.85$~Hz, corresponding to a phonon lifetime of $\tau_{\textrm{coh}}=41.3$~ms. We finally verify the coherence of our electromechanical phononic memory by transferring four microwave coherent states with an $\pi/2$ increment in phase. The state tomography of recovered microwave pulses is presented in Fig.~7\textbf{c}, which shows that the phase of the signal pulse is faithfully recovered after storage.
\bigskip
\\
\noindent\textbf{Coherent energy transfer between distinct high-$Q$ mechanical modes}\\
In the 3C-SiC square membrane-based microwave electromechanical device examined in this study, not only can the (1,3) mode be sideband cooled to its ground state and utilized as a coherent memory, but we can also select other mechanical modes, such as the (3,3) mode. The thermal decoherence rate of the (3,3) mode is $\Gamma_{\textrm{th}}^{3,3}/2\pi=2.35$~Hz, corresponding to a phonon lifetime of $\tau_{\textrm{coh}}=67.75$~ms. Coherent manipulation of phonon transport between distinct mechanical modes is essential for preparing mechanical nonclassical states and for potential applications in bosonic coding with long-lived phononic modes. Since these mechanical modes are optomechanically coupled to the same microwave cavity, they can interact through an intermediary microwave field.

By employing a mechanism similar to STIRAP (stimulated Raman adiabatic passage) in atomic physics, we can couple two distinct mechanical resonators using an effective beam splitter interaction, demonstrating that efficient and coherent state transfer is achievable between these frequency-separated mechanical modes within the same microwave cavity. While energy transfer between two separated mechanical modes has been demonstrated in the classical regime~\cite{Weaver2017Coherent, fedoseev2021stimulated}, in this work, both mechanical modes are initialized in their ground states.

As shown in Figure 8, we illustrate coherent optomechanical state swapping between two frequency-separated mechanical modes in the quantum regime. The state transfer pulse sequence is similar to the scheme depicted in Fig. 1\textbf{e}, but replaces the storage duration with a swapping pulse operation. By repeating the state-swapping pulse sequence for various durations thousands of times, we can calibrate the coherent and thermal components of the phonon number for each mechanical mode (corresponding to each point in Figure 8) using the method demonstrated in Figure 7.
\begin{table*}[h]
    \centering
    \caption{The performance of mechanical resonators in optomechanical (OM) and electromechanical (EM) devices}
    \resizebox{1\textwidth}{!}{ 
        \begin{tabular}{cccccccccc}
            \toprule
            material & \parbox{1cm}{bandgap\\isolation} & structure & frequency & $Q$ factor & total linewidth & pure dephasing & group delay & \parbox{1cm}{ground-state} & coherence time \\
            \midrule
            Al-stairs-drum~\cite{teufel2011circuit} & no & stairs-circular plate (EM) & $2\pi\cdot$10.69~MHz & $3.6\times10^{5}$ & $2\pi\cdot30$~Hz & not report & $\sim5~ms$ & yes & 130~$\mu$s \\
            Al-flat-drum~\cite{Youssefi2023squeezed}  & no & flat circular plate (EM) & $2\pi\cdot$1.8~MHz & ~1.6$\times10^{7}$ & $2\pi\cdot$0.135~Hz & $2\pi\cdot$0.09~Hz & not report & yes & 7.7~ms \\
            SiN~\cite{zhou2013slowing}  & no & double clamped beam (EM) & $2\pi\cdot$1.45~MHz & $6\times10^{5}$ & $2\pi\cdot$14.5~Hz & not report & 3.4~ms & no & not report \\
            SiN~\cite{fink2016quantum}  & yes & phononic crystal beam (EM) & $2\pi\cdot$4.48~MHz & $5.6\times10^{5}$ & $2\pi\cdot$8~Hz & not report & 19.9~ms & yes & not report \\
            SiN~\cite{seis2022ground}  & yes & Lotus membrane (EM) & $2\pi\cdot$1.48~MHz & 5.7$\times10^{8}$ & $2\pi\cdot$2.6~mHz & $2\pi\cdot$0.5~mHz & not report & yes & $\thicksim$100~ms \\
            SiN~\cite{liu2021optomechanical,liu2023coherent} & no & square membrane (EM) & $2\pi\cdot$750~kHz & 7.8$\times10^{7}$ & $2\pi\cdot$9.7~mHz & not report & $\sim$500~s & yes & 55.7~$\mu$s \\
            SiN~\cite{kristensen2024long} & yes & Honeycomb lattice (OM) & $2\pi\cdot$2.4~MHz & $10^{8}$ & $2\pi\cdot$24~mHz & not report & 23~ms & no & 100~$\mu$s \\
            crystalline silicon~\cite{maccabe2020nano}  & yes & phononic crystal beam (OM) & $2\pi\cdot$5~GHz & $\thicksim4.09\times10^{6}$ & $2\pi\cdot$1.22~kHz & $2\pi\cdot$1.219~kHz & not report & yes & 130~$\mu$s\\
            crystalline silicon~\cite{bozkurt2023quantum}  & yes & phononic crystal beam (EM) & $2\pi\cdot$5.08~GHz & $\thicksim2\times10^{5}$ & $2\pi\cdot31.8$~kHz & $2\pi\cdot$30~kHz & not report & yes & 5~$\mu$s \\
            crystalline silicon~\cite{wallucks2020quantum} & yes & phononic crystal beam (OM) & $2\pi\cdot$5.12~GHz & $\thicksim2.58\times10^{6}$ & $2\pi\cdot$1.98~kHz & $2\pi\cdot$1.42~kHz & not report & yes & $\leq112~\mu$s \\
            gallium phosphide (GaP)~\cite{stockill2022ultra} & yes & phononic crystal beam (OM) & $2\pi\cdot$2.81~GHz & $4\times10^{4}$ & $2\pi\cdot$67~kHz & $2\pi\cdot$64.5~kHz & not report & yes & 2.37~$\mu$s \\
            gallium arsenide (GaAs)~\cite{forsch2020microwave} & no & phononic crystal beam (OM) & $2\pi\cdot$2.7~GHz & $1.8\times10^{4}$ & $2\pi\cdot$179~kHz & not report & not report & yes & 0.8~$\mu$s \\
            aluminum nitride (AlN)~\cite{fong2014microwave} & no & microdisk (OM) & $2\pi\cdot$0.78~GHz & $3.86\times10^{3}$ & $2\pi\cdot$202~kHz & not report & 0.76~$\mu$s & no & not report \\
            lithium niobate (LiNbO$_{3}$)~\cite{jiang2019lithium} & yes & phononic crystal beam (OM) & $2\pi\cdot$2.1~GHz & $1.7\times10^{4}$ & $2\pi\cdot$123~kHz & not report & not report & no & not report \\
            crystalline diamond~\cite{lake2021processing} & no & microdisk (OM) & $2\pi\cdot$2.14~GHz & $1.12\times10^{4}$ & $2\pi\cdot$190~kHz & not report & $\thicksim$5~ms & no & not report \\
            3C-SiC~\cite{lu2015high}  & no  & microdisk (OM)  & $2\pi\cdot$579~MHz  & 5.3$\times10^{3}$  & $2\pi\cdot$108~kHz  & not report  & not report  & no  & not report  \\
            3C-SiC~\cite{reigue2023cavity,fogliano2021mapping,mercier2017universal}  & no & nanowire (OM) & $2\pi\cdot$50~kHz & $10^{3}$ & $2\pi\cdot$50~Hz & not clear & not report & no & not report \\
            4H-SiC~\cite{hamelin2019monocrystalline}  & yes & microdisk (OM) & $2\pi\cdot$6.2~MHz & 2.8$\times10^{6}$ & $2\pi\cdot$2.21~Hz & not report & not report & no & not report \\
            3C-SiC~[this work]  & no & square membrane (EM) & \parbox{1.8cm}{$2\pi\cdot$871~kHz\\ mode (1,3)} & 1.2$\times10^{8}$ & $2\pi\cdot$8.2~mHz & $2\pi\cdot$0.28~mHz & 4035~s & yes & 41.3~ms \\
            3C-SiC~[this work]  & no & square membrane (EM) & \parbox{1.8cm}{$2\pi\cdot$1.178~MHz \\ mode (3,3)} & 2$\times10^{8}$ & $2\pi\cdot$6.04~mHz & $2\pi\cdot$0.25~mHz & $>$5000~s & yes & 67.75~ms \\
            \bottomrule
        \end{tabular}
    }
    \par\vspace{0.5em} 
    \textit{Note: the listed $Q$ factors in the table correspond to the total linewidth rather than solely considering energy decay. For these properties of other oscillators not listed in the table, please refer to the reviews~\cite{serra2021silicon,sementilli2022nanomechanical,schmid2016fundamentals,engelsen2024ultrahigh}}
\end{table*}
\bigskip
\\
\noindent\textbf{Discussion}\\
We have constructed a microwave cavity electromechanical system comprising of a 3D microwave cavity and a 3C-phase SiC membrane resonator. We have confirmed that the low-resistance semiconductor 3C-SiC is compatible with superconducting quantum circuits, ensuring a low electromagnetic loss. The non-uniform tensile stress breaks the membrane's rotational symmetry, leading to the splitting of degenerate superposition modes into near-resonant mode pairs with distinct mode shapes. Both modes in the degeneracy-breaking mode pair exhibit a vertical displacement at the exact center, in stark contrast to the situation where only the constructive superposition mode in the degenerate mode pair has a central vertical displacement, thereby increasing the mechanical modes that the microwave cavity can read. By combining the notch-type asymmetric structure design of the bottom electrode with the breaking of mechanical modes degeneracy, the shared microwave cavity field enables the electromechanical readout of 21 mechanical modes. Additionally, we found that the extremely high thermal conductivity of the 3C-SiC membrane is another advantage for mitigating the notorious mechanical heating effect observed in other optomechanical or electromechanical devices. The 3C-phase SiC membrane chip effectively thermalizes to the operating temperature of the refrigerator, resulting in a four orders of magnitude improvement in the $Q$-factors of the mechanical modes compared to the performance measured at room temperature. Nineteen of the detected mechanical modes (over 90\%) demonstrate $Q$-factors exceeding $10^{8}$. Frequency instability and pure dephasing are increasingly prominent limitations for further improving the performance of high $Q$ mechanical resonators. To better highlight the advantages of using a 3C-SiC square membrane as a mechanical resonator in microwave electromechanical devices, the performance of mechanical resonators made from other materials, such as crystalline silicon or amorphous SiN, is summarized in Table 2.

The table illustrates that previously, SiC films were primarily utilized in conjunction with optical microcavities to facilitate cavity optomechanical coupling. Our work introduces the application of crystalline SiC membrane in the realm of microwave electromechanical (EM) coupling. Unlike other single-crystal materials, e.g., crystalline silicon nanobeam~\cite{maccabe2020nano,bozkurt2023quantum,wallucks2020quantum}, which exhibit extremely low energy decay rate (with a quality factor \( Q \sim 10^{10}\)), the frequency instability and jitter at low temperatures introduces additional decoherence. This results in a coherence lifetime for the oscillator on the order of hundreds of microseconds ($\tau_{\textrm{coh}}\sim100~\mu$s). In contrast, our research demonstrates the exceptional frequency stability of 3C-SiC oscillators at low temperatures, with coherence times approach to the order of a hundred of milliseconds, which is over three orders of magnitude greater than that of other single-crystal oscillators. This enhanced frequency stability allows the microwave field group delay induced by 3C-SiC square membrane oscillators to exceed one hour, surpassing the best reported levels of hundreds of seconds in SiN membrane oscillator systems~\cite{liu2021optomechanical,liu2023coherent}.

As a perspective, integrating phonon bandgap engineering and dissipative dilution techniques will further enhance the quality factor of 3C-SiC oscillators. Replacing the 3D aluminum cavity with an annealed copper or on-chip 2D superconducting LC resonator can reduce the cavity field heating effect, enabling a lower phonon occupancy number. Defects in 3C-SiC, such as silicon vacancies (VSi) and carbon vacancies (VC), can serve as qubits with long coherence times. Optical fields can be used to manipulate and detect the quantum states of these single spins. By integrating high-$Q$ and long-lived 3C-SiC mechanical membrane resonators, a functional interface can be established to transfer information between disparate quantum elements, including microwave photons, superconducting qubits, defect spins, and even optical photons. By combining disparate quantum elements through 3C-SiC mechanical resonators, we can enhance functionality and expand the toolbox for quantum information processing~\cite{barzanjeh2022optomechanics,chu2020perspective,XU2022100016Hybrid}. Exploring these avenues holds great potential to enable coherent information transfer and advance hybrid quantum architectures with enhanced capabilities. 
\bigskip
\\
\noindent\textbf{Methods}\\
\noindent\textbf{Device fabrication}\\
The device consists of a 3D aluminum (Al) superconducting cavity and a mechanical capacitor chip. The Al-rectangle cavity is shaped and polished after machining. The mechanical capacitor chip is assembled by using a flip-chip bonder. The upper parallel-plate electrode is made on a 3C-phase silicon carbide membrane with 400~$\mu$m thick silicon frame (Norcada PSCX5050A). The SiC window was metalized with a circular Al electrode deposited by electron beam evaporation (JEB-4, Adnanotek). Commercial ACC$\mu$RA flip-chip bonder was used to align the center of the upper layer electrode on the squared SiC window to the center of the bottom parallel plate electrode of the coupling antenna chip. The main structure of the bottom antenna chip is an H-shaped electrode with 120~nm thick niobium sputtered on a high resistance silicon (orientation of $\langle100\rangle$) substrate with a thickness of 500 $\mu$m. The Nb thin film is sputtered through standard physical vapor deposition technology (Syskey Technology, SP-lC4-A06). The sputter working parameters are $5\times 10^{-3}$~hPa, power 150~W and time 800~s, respectively. To mitigate the influence of oxidation on the resonator performance, the Nb film was placed in nitrogen (a pressure of 1000~Pa) for 35 minutes. The Nb/Si wafer was then subjected to the lithography processes. We selected S1813 photoresist as the coating layer with a spinning rate of 3000 rpm and an operating time of 1 minute. The photoresist was then baked at 115~$^{\circ}\rm C$ for 2 minutes. We used the DWL 66+ laser lithography tool (Heidelberg Instruments) to transfer the pattern. In the process of laser direct-writing, the laser power, the intensity, the filter, and the focus were set at 70~mW, 30\%, 12.5\%, and -20\%, respectively. The development process was carried out at 1 minute in the MF-319 developer, followed by rinsing in DIW fixing solution. When the pattern was completely transferred, the unwanted Nb film was removed in the RIE process (RIE-10NR, Samco Inc).
\bigskip
\\
\noindent\textbf{Measurement methods}\\
The device was mounted on the 10~mK mixing-chamber plate of a dilution refrigerator and a circulator was used for reflection measurements. The cancellation line is combined with the output line via a directional coupler. By adjusting the amplitude and phase of the canceling tone, the reflected pump-tone can be effectively canceled to avoid the HEMT saturation.
An arbitrary waveform generator (Tektronix AWG5014C) with a mixer is used to shape the sideband pump-tone (with frequency $\omega_{\textrm{c}}-\omega_{\textrm{m}}$), amplify-tone (with frequency $\omega_{\textrm{c}}+\omega_{\textrm{m}}$) and signal-tone (with frequency $\omega_{\textrm{c}}$) to pulse. These three tones are combined through a splitter at room temperature and then transmitted down to the device. To increase the on-off switching ratio of the pulses, we have used high-speed microwave switchers after the mixer for each tone, respectively. The control signal of the switch and the pulse signal of the mixer are simultaneously generated by the AWG in the marker channel and analog channel. A real-time signal analyzer (Tektronix RSA5126B) is used for digitalizing the output voltage signal. For the state tomography in the time domain, the frequency center of local RSA acquiring is slightly shifted from signal-tone (i.e., fixed at $\omega_{\textrm{c}}/2\pi-400$~Hz), to make the acquired IQ signal have a small oscillation frequency. The storage time from write-in to read-out is controlled by the AWG in the time domain. To synchronize the RSA signal acquisition with a signal pulse, the RSA is triggered by an AWG pulse. For device $S_{21}$ parameter measurement, the PNA (Agilent N5232A) probe tone is also combined into the input line. When taking the $S$ parameter measurement and spectrum measurement, the amplify-tone and signal-tone are switched off, and the pump-tone is always kept on. All instruments are phase-locked by a 10~MHz rubidium frequency standard.
\bigskip
\\
\noindent\textbf{Data Availability}\\
All data supporting the findings of this study are available within the article and the Supplementary Information file, or available from the corresponding authors upon request.

\bigskip
\noindent\textbf{Acknowledgements}\\
This work is supported by the National Natural Science Foundation of China (Grants No.~92365210). Y.~Liu acknowledges the support of Beijing Municipal Science and Technology Commission (Grant No.~Z221100002722011), Beijing Natural Science Foundation (Z240007), and Young Elite Scientists Sponsorship Program by CAST (Grant No.~2023QNRC001). This work is also supported by the National Key Research and Development Program of China (Grant No.~2022YFA1405200) and the National Natural Science Foundation of China (No.~12374325, No.~62074091, No.~12304387). We acknowledge the facilities and technical support of Otaniemi research infrastructure for Micro and Nanotechnologies (OtaNano). The work was supported by European Research Council (101019712). The work was performed as part of the Research Council of Finland Centre of Excellence program (project 336810). We acknowledge funding from the European Union's Horizon 2020 research and innovation program under the QuantERA II Programme (13352189).
\bigskip
\\
\noindent\textbf{Author Contributions}\\
Y.~L. carried out the device design, fabrication and drafted the manuscript. H.~W. metalizes the SiC membrane. H. S. conducts the room temperature Doppler measurements. Q.~L., Y.~L., and T.~L. developed the measurement schemes. Y.~L. and M.~A.~S. conceived the project and coordinated the research. All authors analyzed the measurement results and gave their final approval for publication.
\bigskip
\\
\noindent\textbf{COMPETING INTERESTS}\\
The authors declare no competing interests.
\end{document}